\documentclass[english,10pt,aps,pra,twocolumn,superscriptaddress,floatfix,fleqn]{revtex4-1}
\pdfoutput=1
\usepackage[utf8]{inputenc}
\usepackage[T1]{fontenc}
\usepackage{amsmath}
\usepackage{amssymb}
\usepackage{amsfonts}
\usepackage{bm}
\usepackage{bbm}
\usepackage{epsfig}
\usepackage{grffile}
\usepackage{amsthm}

\usepackage[usenames,dvipsnames]{color}
\definecolor{dblue}{rgb}{0,0.1,.6}
\definecolor{dlblue}{rgb}{0,0.15,.75}
\definecolor{dred}{rgb}{.6,0.1,0}

\newcommand{\dom}[1]  {\textcolor{dred}{#1}}
\newcommand{\jon}[1]  {\textcolor{dlblue}{#1}}

\usepackage[colorlinks=true,citecolor=dblue,linkcolor=dblue,urlcolor=dblue]{hyperref}
\usepackage[all]{hypcap}

\usepackage{algorithm}
\usepackage{algpseudocode}

\newcommand{\id}{\mathbbm{1}}
\newcommand{\bra}{\langle}
\newcommand{\ket}{\rangle}
\newcommand{\pdag}{{\phantom{\dag\!}}}
\newcommand{\Tr}{\operatorname{Tr}}
\newcommand{\mc}[1]{\mathcal{#1}}
\renewcommand{\vec}[1]{{\boldsymbol{#1}}}

\newcommand{\C}{\mc{C}}
\newcommand{\E}{\mc{M}}
\renewcommand{\O}{\mc{O}}
\renewcommand{\S}{\mc{S}}
\newcommand{\T}{\mc{T}}

\newcommand{\hH}{{H}}
\newcommand{\hh}{{h}}
\newcommand{\dm}{{\rho}}
\renewcommand{\vr}{\vec{r}}

\newcommand{\CC}{\mathbb{C}}

\newcommand{\lin}[2]{\xrightarrow{(#1,#2)}}
\newcommand{\tq}{$\vdots$}
\newcommand{\tqs}{{$\vdots$}}
\newcommand{\TL} {\text{TL}}
\newcommand{\TC} {\text{TC}}
\newcommand{\TR} {\text{TR}}
\newcommand{\ML} {\text{ML}}
\newcommand{\MC} {\text{MC}}
\newcommand{\MR} {\text{MR}}
\newcommand{\BL} {\text{BL}}
\newcommand{\BC} {\text{BC}}
\newcommand{\BR} {\text{BR}}
\newcommand{\tL} {\text{L}}
\newcommand{\tR} {\text{R}}
\newcommand{\tT} {\text{T}}
\newcommand{\tB} {\text{B}}
\newcommand{\tC} {\text{C}}

\newcommand{\Emph}[1]{\emph{#1.} ---}

\newtheorem{lemma}{Lemma}

\renewcommand{\thesection}{\Roman{section}}
\renewcommand{\thesubsection}{\thesection.\arabic{subsection}}
\makeatletter
\renewcommand{\p@subsection}{}
\renewcommand{\p@subsubsection}{}
\makeatother

\newcommand{\duke}  {Department of Physics, Duke University, Durham, North Carolina 27708, USA}
\newcommand{\dqc}   {Duke Quantum Center, Duke University, Durham, North Carolina 27701, USA}
\newcommand{\tz}    {Tensor Center, Auf dem Dresch 15, 52152 Simmerath, Germany}

\newcommand{\Title} {\texorpdfstring{Scaling of contraction costs for entanglement renormalization algorithms\\ including tensor Trotterization and variational Monte Carlo}{Scaling of contraction costs for entanglement renormalization algorithms including tensor Trotterization and variational Monte Carlo}}
\newcommand{\Authors}
{
\author{Thomas Barthel}
\affiliation{\duke}
\affiliation{\dqc}
\affiliation{\tz}
\author{Qiang Miao}
\affiliation{\dqc}
}
\newcommand{\Date} {July 27, 2024}

\begin{document}

\title{\Title}
\Authors

\date{\Date}

\begin{abstract}
The multi-scale entanglement renormalization ansatz (MERA) is a hierarchical class of tensor network states motivated by the real-space renormalization group. It is used to simulate strongly correlated quantum many-body systems. For prominent MERA structures in one and two spatial dimensions and different optimization strategies, we determine the optimal scaling of contraction costs as well as corresponding contraction sequences and algorithmic phase diagrams. This is motivated by recent efforts to employ MERA in hybrid quantum-classical algorithms, where the MERA tensors are Trotterized, i.e., chosen as circuits of quantum gates, and observables as well as energy gradients are evaluated by sampling causal-cone states. We investigate whether tensor Trotterization and/or variational Monte Carlo (VMC) sampling can lead to quantum-inspired classical MERA algorithms that perform better than the traditional optimization of full MERA based on the exact evaluation of energy gradients. Algorithmic phase diagrams indicate the best MERA method depending on the scaling of the energy accuracy and the optimal number of Trotter steps with the bond dimension. The results suggest substantial gains due to VMC for two-dimensional systems.
\end{abstract}

\maketitle

\section{Introduction}\label{sec:intro}
Tensor network states were initially developed in condensed matter physics to study thermal states of classical systems and as a variational method for quantum ground states \cite{Karmers1941-60,Baxter1968-9,Nightingale1986-33,Fannes1992-144,White1992-11,Niggemann1997-104,Nishino2000-575,Verstraete2004-7,Vidal-2005-12,Schollwoeck2011-326,Orus2014-349,Cirac2021-65}. In the realm of quantum many-body physics, they are also used for the investigation of finite-temperature states, equilibrium and non-equilibrium dynamics, and driven-dissipative systems. More recently, tensor network methods have, for example, been adapted for machine learning \cite{Cohen2016-29,Stoudenmire2016-29,Novikov2016_05,Stoudenmire2018-3,Grant2018-4,Liu2019-21,Huggins2019-4,Cheng2021-103,Vieijra2022_02,Chen2024-46} and the investigation of stochastic dynamics in classical network systems \cite{Barthel2018-97,Barthel2020-1,Crotti2023-120,Crotti2024_11}.

The multi-scale entanglement renormalization ansatz (MERA) as introduced by Vidal \cite{Vidal-2005-12,Vidal2006} is a hierarchical class of tensor network states motivated by the real-space renormalization group. It is particularly suitable for the representation of ground states of critical one-dimensional (1D) systems. In contrast to matrix product states (MPS) \cite{Fannes1992-144,White1992-11,Rommer1997,PerezGarcia2007-7,Schollwoeck2011-326}, MERA bond dimensions need not be increased with increasing system size, and MERA make it easy to extract critical properties like operator-scaling dimensions for 1D systems
\cite{Giovannetti2008-101,Pfeifer2009-79,Evenbly2010-82}. The situation changes in $D\geq 2$ spatial dimensions. There, MERA form a subclass of the projected entangled-pair states (PEPS). In particular, each $D\geq 2$ MERA can be mapped to a PEPS with system-size independent bond dimension. So, for fixed bond dimension $\chi$, $D\geq 2$ MERA  obey the entanglement-entropy area law \cite{Barthel2010-105}, and one would need the computationally more expensive class of branching MERA \cite{Evenbly2014-112} to capture, e.g., log-area law entanglement. Compared to PEPS, MERA have the advantage of a causal structure (causal cone) which makes it possible to efficiently evaluate expectation values of local observables without approximations. Also, despite having loops, MERA form closed sets \cite{Barthel2022-112,Landsberg2012-12}.
Compared to MPS, MERA have a much more suitable entanglement structure for systems in $D\geq 2$ dimensions. The causal structure of MERA and the fact that their tensors are by construction isometric leads to promising approaches for the simulation of quantum many-body systems on quantum computers \cite{Kim2017_11,Miao2021_08,Haghshenas2022-12,Miao2023_03,Barthel2023_03,Miao2024-109,Haghshenas2023_05,Job2024_04,Miao2024_12}
\begin{figure*}[t]
	\includegraphics[width=0.96\textwidth]{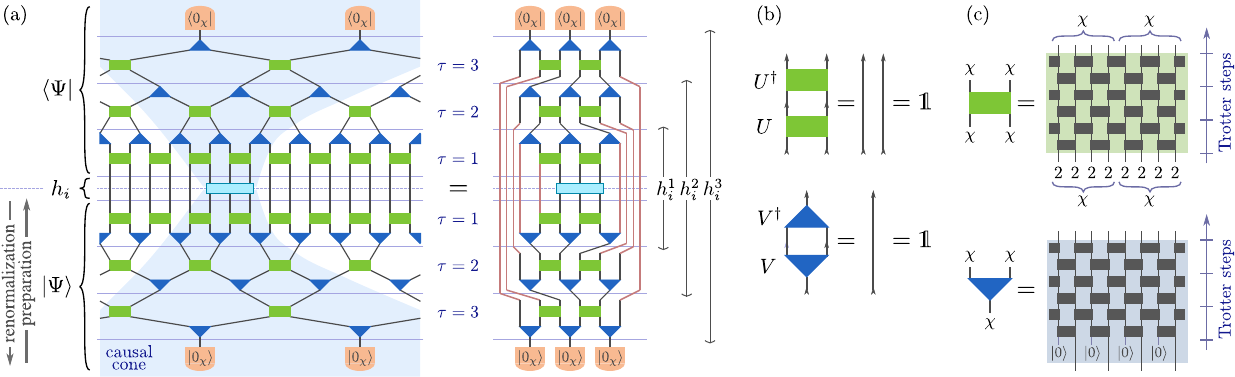}
	\caption{\label{fig:MERA-Trotter}\textbf{MERA expectation values and Trotterized tensors}.
	(a,b) The MERA $|\Psi\ket$ consists of disentanglers (boxes) and isometries (triangles). Contraction lines between the tensors correspond to renormalized-site vector spaces with bond dimension $\chi$. Due to the isometry constraints, the expectation value $\bra\Psi|\hh_i|\Psi\ket$ of the three-site operator $\hh_i$ only depends on tensors in its causal cone. For the 1D binary MERA, the cone has width three, i.e., contains at most three renormalized sites in each layer.
	(c) We also consider Trotterized tensors as suggested in the context of variational quantum algorithms \cite{Kim2017_11,Miao2021_08,Haghshenas2022-12,Miao2023_03,Barthel2023_03,Miao2024-109,Haghshenas2023_05,Job2024_04,Miao2024_12}, where each tensor is a circuit of (multi-qubit) quantum gates. Specifically, the figure shows brickwall circuits of two-qubit gates comprising $t=3$ Trotter steps, and each renormalized site corresponds to four qubits ($\chi=2^4=16$).}
\end{figure*}

\Emph{Content} In this work, we investigate whether computation costs for MERA simulations can be reduced
\begin{enumerate}\itemsep0pt
 \item[(a)] by sampling pure causal-cone states, i.e., applying variational Monte Carlo (VMC) \cite{Ferris2012-85,Sandvik2007-99,Schuch2008-100} instead of the usual optimization based on exact energy gradients (EEG) \cite{Evenbly2009-79},
 \item[(b)] by constraining the MERA tensors to be circuits of quantum gates with a certain depth (typically brickwall circuits of two-qubit gates), which we refer to as \emph{Trotterized MERA} (TMERA) \cite{Miao2021_08}, or
 \item[(c)] by a combination of both.
\end{enumerate}
We determine how the computation costs for the four resulting algorithms (full-MERA EEG, full-MERA VMC, TMERA EEG, and TMERA VMC) scale in the bond dimension $\chi$, specify corresponding tensor contraction sequences, and provide algorithmic phase diagrams in a comprehensive analysis of the six most prominent types of 1D and 2D MERA.

\Emph{Trotterized MERA} MERA tensor networks consist of unitary disentanglers and projections (isometries) that remove unimportant local degrees of freedom \cite{Vidal-2005-12,Vidal2006}. We will often refer to MERA with unconstrained tensors of bond dimension $\chi$ as \emph{full MERA} (fMERA).
For the implementation of MERA on quantum computers, one can constrain the tensors and realize them as brickwall circuits of two-qubit gates \cite{Kim2017_11,Miao2021_08,Haghshenas2022-12,Miao2023_03} as shown in Fig.~\ref{fig:MERA-Trotter}c. We refer to the depth of these circuits as the number of \emph{Trotter steps} $t$. With increasing $t$, the two-qubit gates get closer and closer to identities \cite{Miao2023_03}, and the circuits are very similar to those obtained in Lie-Trotter-Suzuki decompositions for time evolution operators \cite{Trotter1959,Suzuki1976-51,Barthel2020-418,Childs2021-11}. This motivates the term ``Trotterization''. The ``deep MERA'' of Refs.~\cite{Kim2017_11,Job2024_04} are closely related and can be interpreted as a class of TMERA with a specific linear relation between the number of Trotter steps $t$ and the width $\sim 2t$ of the causal cones. For the purpose of this work, the tensors in a TMERA may, more generally, be circuits of $(k<K)$-qubit gates organized in $t$ layers, where the bound $K$ is independent of the bond dimension $\chi$, and the number of gates in each Trotter step scales linearly in the number of input qubits. In the spirit of quantum-inspired algorithms, we analyze here whether \emph{classical} TMERA can outperform fMERA simulations.

\begin{figure*}[t]
	\includegraphics[width=1\textwidth]{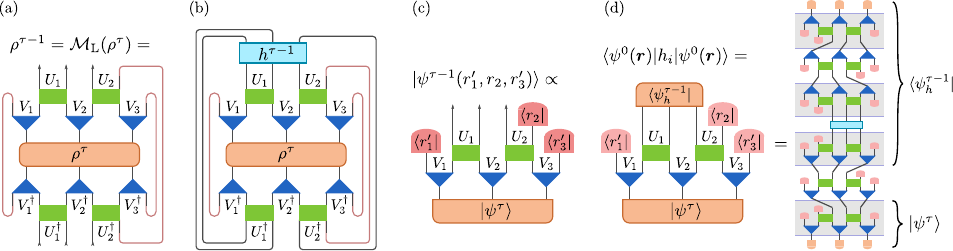}
	\caption{\label{fig:1DbinaryMERA}\textbf{Optimization of 1D binary MERA.}
	The optimization of MERA requires the evaluation of environment tensors. (a) In the EEG approach, we propagate interaction terms $h^\tau$ in the renormalization direction and causal-cone density operators $\dm^\tau$ in the preparation direction according to Eq.~\eqref{eq:prop-EEG}, shown here for the left-moving transition $\E_\tL$. (b) Environment tensors (the EEG) are obtained by removing one of the disentanglers $U_k$ or isometries $V_k$ from the shown tensor network for $\bra\Psi|\hh_i|\Psi\ket=\Tr\big[\hh_i^{\tau-i}\E_\tL(\dm_i^\tau)\big]$.
	(c) The VMC algorithm is based on sampling pure causal-cone states \eqref{eq:prop-VMC}. Instead of tracing out the renormalized sites that leave the causal cone, one applies projective measurements with suitable measurement bases $\{|r_k\ket\}$. (d) During the importance sampling, energy gradients (environment tensors) are evaluated stochastically by removing one of the MERA tensors from energy-expectation networks like the one shown. The lighter color for the projections $\bra r_k|$ indicate that the measurement results and associated probabilities are already known in this step.}
\end{figure*}

\Emph{EEG versus VMC for MERA} In the conventional EEG optimization of MERA \cite{Evenbly2009-79}, one cycles through all tensors, minimizing in each step the energy expectation value $E=\bra\Psi|\hH|\Psi\ket$ with respect to the elements of one MERA tensor, where we assume lattice Hamiltonians $\hH=\sum_i\hh_i$ with finite-range interaction terms $\hh_i$. For the optimization with respect to a tensor $U_k$, one computes the corresponding \emph{environment tensor} $\partial_{U_k}E$ exactly, which is obtained by removing $U_k$ from the tensor network that gives the expectation value $E$. This environment tensor is nothing but the exact energy gradient (EEG). Central steps are to propagate Hamiltonian terms $\hh^{\tau}_i=\E^\dag_{\tau,i}(\hh^{\tau-1}_i)$ with $\hh^0_i\equiv\hh_i$ in the renormalization direction and causal-cone density operators 
\begin{equation}\label{eq:prop-EEG}
	\dm^{\tau-1}_i=\E_{\tau,i}(\dm^\tau_i)
\end{equation}
in the preparation direction of the MERA. The latter are mixed states for the renormalized sites in the causal cone at MERA layer $\tau$. $\E_{\tau,i}$ are the layer transition maps which consist in applying the isometries and disentanglers of layer $\tau$ and tracing out the renormalized sites that leave the causal cone. See, for example, Refs.~\cite{Evenbly2009-79,Barthel2023_03}. In the case of the 1D binary MERA shown in Figs.~\ref{fig:MERA-Trotter} and \ref{fig:1DbinaryMERA}, we have for example
\begin{multline}\label{eq:prop-EEG-example}
	 \dm^{\tau-1}=\Tr_{1,5,6}\big[(\id_\chi\otimes U_1\otimes U_2\otimes \id_\chi)(V_1\otimes V_2\otimes V_3)\,\dm^\tau\\
	 \times(V_1\otimes V_2\otimes V_3)^\dag(\id_\chi\otimes U_1\otimes U_2\otimes \id_\chi)^\dag\big].
\end{multline}
In contrast, the VMC approach applies \emph{importance sampling}, where pure causal-cone states $|\psi_i^\tau\ket$ are propagated in the preparation direction, sites that leave the causal cone are measured protectively, and we sample over the measurement results according to their Born-rule probabilities \cite{Ferris2012-85}. For example, the VMC layer transition of the 1D binary MERA shown in Fig.~\ref{fig:1DbinaryMERA}c is
\begin{multline}\label{eq:prop-VMC}
	 |\psi^{\tau-1}(r'_1,r_2,r'_3)\ket\propto \big(\bra r'_1|\otimes\id_{\chi^3}\otimes\bra r_2|\otimes\bra r'_3|\big)\\
	 \times(\id_\chi\otimes U_1\otimes U_2\otimes \id_\chi)(V_1\otimes V_2\otimes V_3)\,|\psi^\tau\ket,
\end{multline}
where $r'_1,r_2,r'_3$ are the results for measurements on sites $1,5,6$ which leave the causal cone (sites $2,3,4$ remain). Correspondingly, environment tensors are evaluated stochastically.

In the following $\epsilon:=(E-E_0)/N$ denotes the accuracy of the MERA energy density $E/N$. For critical systems, it improves polynomially with increasing bond dimension. We characterize the convergence 
\begin{equation}\label{eq:eps-chi}
	\epsilon \sim \chi^{-\beta}
\end{equation}
by the critical exponent $\beta$ which depends on the model (the system's universality class) and MERA type. Similarly, for TMERA, the optimal number of Trotter steps increases according to a power law
\begin{equation}\label{eq:t-chi}
	t \sim \chi^{p}.
\end{equation}
After the determination of optimal tensor contraction sequences for the different algorithms, we can deduce algorithmic phase diagrams, spanned by the model-dependent exponents $\beta$ and $p$. For several critical 1D systems, we determine the exponents numerically.

\Emph{Optimized tensor contraction sequences} 
The EEG and VMC algorithms are based on the contraction of tensor networks as shown in Fig.~\ref{fig:1DbinaryMERA} -- in particular, those of the energy-environment tensors. Following the discussion in Refs.~\cite{Lam1997-07,Hartono2005-155,Pfeifer2014-90}, we solve the associated \emph{single-term optimization} problems \cite{Hartono2005-155} through breadth-first constructive search: For a given tensor network, we determine all subgraphs that can be contracted at cost $\O(\chi^{\alpha_1})$ and corresponding contraction sequences, then all the subgraphs that can be contracted with some cost $\O(\chi^{\alpha_2>\alpha_1})$, and iteratively increase $\alpha_i$ until finding an optimal sequence that contracts the entire network.

The contraction problem for the importance-sampling step of VMC is complicated by the need to account for the projective measurements. The sequence optimization costs are however relatively low such that we can employ a depth-first search with cost-based pruning.

\section{Full-MERA contraction costs}\label{sec:fMERA}

\subsection{Contractions for the EEG approach}\label{sec:fMERA-EEG}
The problem of finding optimal tensor-network contraction sequences is NP-complete \cite{Lam1997-07}. Several exhaustive approaches determining optimal sequences \cite{Lam1997-07,Hartono2005-155,Pfeifer2014-90,Liang2021-15} as well as heuristic approaches finding good sequences \cite{Schindler2020-1,Schutski2020-102,Jermyn2020-8,Huang2021-1,Gray2021-5} have been developed.

The relevant tensor networks considered here are connected, and all bond dimensions (tensor dimensions) are integer powers $\chi^\nu$ of a base dimension $\chi$. Furthermore, we regard a contraction sequence as \emph{optimal} if the leading term of the associated cost $\O(\chi^\alpha)$ has the smallest possible exponent $\alpha$. So, we are not concerned with optimizing the coefficient of the leading term or subleading terms, which simplifies matters.

Let us summarize basic facts about optimal tensor network contractions for the case of unconstrained tensors with proofs provided in Appx.~\ref{appx:TNbasics}:
\begin{lemma}\label{lemma1}
  In the optimal contraction of two tensors, we can always sum simultaneously over all common indices.
\end{lemma}
\begin{lemma}\label{lemma2}
  Every tensor network can be contracted optimally through pairwise contractions, i.e., by contracting exactly two tensors in each step.
\end{lemma}
\begin{lemma}\label{lemma3}
  For optimal contraction sequences, we never need to consider outer products of tensors, i.e., no contractions of tensors without shared indices.
\end{lemma}

To determine optimal contraction sequences for the fMERA layer-transition and energy-environment tensor networks occurring in the EEG algorithm, one can follow a \emph{depth-first search} approach, suggesting in each step the contraction of two connected tensors and continuing with further contractions as long as the contraction costs do not exceed a certain threshold. The search is stopped when a complete contraction with costs below a predefined threshold is identified.
The number of possible sequences roughly increases exponentially with the number of tensors in the network, which is reflected in the NP-completeness of the sequence optimization problem \cite{Lam1997-07}. One can speed up the search by cost-based pruning of the search tree and by creating look-up tables for previously encountered partially contracted networks. However, we found that such a depth-first search is only efficient enough for 1D MERA and for the VMC importance-sampling step in 2D; it is generally too slow for the remaining contraction steps for 2D MERA.
\begin{figure}[t]
\hrule
\vspace{0.5em}
\begin{algorithmic}[1]
\State $\alpha\gets 1$, $\alpha'\gets \infty$ 
\State $\S_1\gets\{A_1,\dotsc,A_n\}$, $\S_2\gets \{\},\ \dotsc,\ \S_n\gets \{\}$
\While{$|\S_n|==0$} 
  \For{$m=2,\dotsc,n$; \ $k=1,\dotsc,\lfloor m/2\rfloor$} 
      \ForAll{$T\in \S_k$, $T'\in \S_{m-k}$}
        \If{\Call{contractible}{$T,T'$} \textbf{and} $T\,T'\notin\S_m$}
          \State $\alpha'' \gets$ \Call{costExponent}{$T,T'$}
          \If{$\alpha''==\alpha$}
          \State $\S_m \gets \S_m \cup \{T\,T'\}$
          \State Also store contraction seq.\ and cost.
          \Else
          \State $\alpha' \gets \min(\alpha',\alpha'')$
          \EndIf
        \EndIf
      \EndFor
  \EndFor
  \State $\alpha \gets \alpha'$
\EndWhile 
\end{algorithmic}
\vspace{0.4em}
\hrule
\caption{\label{alg:breadthFirst}\textbf{Breadth-first constructive search for contraction sequences}. To optimize MERA, we need to contract tensor networks for energy environments as shown in Fig.~\ref{fig:1DbinaryMERA}. Given such a network with elementary tensors $A_1,\dotsc,A_n$ the constructive search algorithm finds optimal contraction sequences. The function \textproc{contractible}$(T,T')$, checks whether the tensors (subnetworks) $T$ and $T'$ share none of the elementary tensors $A_i$ and whether they are connected by a contraction edge (common index). The function \textproc{costExponent}{$(T,T')$}, returns the exponent $\gamma$ for the leading term in the contraction cost $\O(\chi^{\gamma})$ for $T\,T'$. Sections~\ref{sec:fMERA-VMC} and \ref{sec:TMERA-costOpt} describe minor modifications for TMERA and VMC importance sampling.}
\end{figure}

Following the discussion in Refs.~\cite{Lam1997-07,Hartono2005-155,Pfeifer2014-90}, we employ a more efficient \emph{breadth-first constructive search}. Let the given tensor network consist of the $n$ elementary tensors collected in the set $\S_1=\{A_1,\dotsc,A_n\}$. Now let $\S_m(\alpha)$ denote the set of all tensors that can be obtained at cost $\O(\chi^\alpha)$ by contracting $m\leq n$ of the elementary tensors of the tensor network. The algorithm iteratively increases $\alpha$ starting from $\alpha=1$. Say, we have $\S_1(\alpha)\equiv\S_1,\S_2(\alpha),\dotsc,\S_n(\alpha)$, and let $\alpha'>\alpha$ be the smallest real number such that there exist subgraphs of the tensor network that can be contracted with cost $\O(\chi^{\alpha'})$ and are not yet contained in $\bigcup_m\S_m(\alpha)$. Now, we build the sets $\S_2(\alpha'),\dotsc,\S_n(\alpha')$ by initializing them with all elements contained in the sets $\S_m(\alpha)$ and adding the missing elements as follows. For $\S_m(\alpha')$ and all $k=1,\dotsc, m-1$, find all tensors $T\in\S_k(\alpha)$ and $T'\in\S_{m-k}$ such that the sub-networks corresponding to $T$ and $T'$ share no vertices (share no elementary tensors $A_i$), are connected through at least one edge of the full tensor network, and can be contracted at cost $\O(\chi^{\alpha'})$. If these conditions are obeyed, add the contraction $T\,T'$ to $\S_m(\alpha')$; also store the corresponding contraction sequence and cost. This process is continued until $\S_n(\alpha')$ is non-empty, i.e., until we have found an optimal contraction sequence for the full network. This algorithm is summarized in Fig.~\ref{alg:breadthFirst}.

\subsection{Contractions for the VMC approach}\label{sec:fMERA-VMC}
The VMC approach for MERA, as discussed by Ferris and Vidal in Ref.~\cite{Ferris2012-85} and in the introduction (Sec.~\ref{sec:intro}), stochastically propagates pure states for the renormalized sites in the causal cone in the preparation direction,
\begin{equation}\label{eq:VMC-prop}
	|\psi_i^\tau\ket\to|\psi_i^{\tau-1}\ket
\end{equation}
as in Eq.~\eqref{eq:prop-VMC}. In contrast to the layer transitions in the EEG optimization \eqref{eq:prop-EEG}, sites that leave the causal cone are not traced out (which would lead to mixed states) but are measured projectively. For a left-moving layer transition in the 1D binary MERA shown in Fig.~\ref{fig:1DbinaryMERA}c, this means that we start with three renormalized sites, apply one isometry to each of them, yielding six sites, apply two disentanglers on the four central sites, and finally apply a projective measurement for the left-most site and the two right-most sites; see Eq.~\eqref{eq:prop-VMC}. Let us collect in a vector $\vr$ all measurement outcomes for the complete propagation $\tau=T\to T-1\to\dotsc\to 0$ from the final layer to the physical layer, and let us denote the associated Born-rule probability by $p_\vr$. The expectation value of a local interaction term is then
\begin{equation}\label{eq:VMC-expect}
	\bra\Psi|\hh_i|\Psi\ket = \sum_\vr p_\vr \bra\psi_i^{0}(\vr)|\hh_i|\psi_i^{0}(\vr)\ket,
\end{equation}
where $|\Psi\ket$ is the MERA, and $|\psi_i^{0}(\vr)\ket$ is the normalized causal-cone state for the support of $\hh_i$ and measurement results $\vr$.

For the tensor contractions occurring in the importance-sampling layer propagation \eqref{eq:VMC-prop}, we need to make sure that we have a valid (causal-cone) quantum state, before doing any measurements. In principle, this means that some measurements of renormalized sites that leave the causal cone cannot be done immediately after the contraction of the corresponding tensor. In fact, we encountered sequences where some intermediate contraction results are not quantum states [cf.\ Eq.~\eqref{eq:opt2D2x2to1-TMERAimp-a}], and sequences with postponed measurements [cf.\ the second sequence in Eq.~\eqref{eq:opt2D2s3x3to1-imp}].

The projective measurements can be done in any complete (orthonormal) basis. We analyze two cases:
\begin{enumerate}
 \item[(a)] Measuring in an \emph{eigenstate measurement basis} (EMB): Consider that we just contracted tensor $U_k$ and that we want to measure $n$ of the renormalized sites of its image that leave the causal cone (each site associated with a $\chi$-dimensional bond vector space). 
 Let $\varrho_n=\sum_{r=1}^{\chi^n}p_r|r\ket\bra r|$ denote the corresponding $n$-site reduced density operator with eigenstates $|r\ket$ and eigenvalues $p_1\geq p_2\geq \dotsc\geq 0$.
 Then, the EMB is the orthonormal basis
 \begin{equation}\label{eq:EMB}
  	\{|r\ket\,|\,r=1,\dotsc,R\leq \chi^n\}
 \end{equation}
 consisting of the eigenstates with $p_r> 0$ \cite{Ferris2012-85}.
 \item[(b)] Measuring in the \emph{computational measurement basis} (CMB): The MERA tensors are stored in a certain basis. One can employ this basis for the projective measurements. This can reduce computation costs for some steps because the projection then simply corresponds to fixing one tensor index to a certain value (tensor slicing).
\end{enumerate}

As in the more traditional EEG approach, the energy optimization can be done by using Riemannian gradient descent or quasi-Newton methods as described in Refs.~\cite{Hauru2021-10,Luchnikov2021-23,Miao2021_08} or by using the original fixed-point iteration based on singular value decompositions of environment tensors \cite{Evenbly2009-79}. It has been pointed out that the latter method might be less stable in the VMC approach \cite{Ferris2012-85}.
In any case, we need to sample energy derivatives to sufficient precision. If $U_k$ is a tensor in the causal cone of the interaction term $\hh_i$, then the tensor networks for the gradient
\begin{equation}\label{eq:VMC-grad}
	\partial_{U_k}\bra\Psi|\hh_i|\Psi\ket = \sum_\vr p_\vr \partial_{U_k}\bra\psi_i^{0}(\vr)|\hh_i|\psi_i^{0}(\vr)\ket
\end{equation}
are given by those of the expectation value \eqref{eq:VMC-expect} with $U_k$ removed ($U_k^\dag$ remains). These tensor networks, as illustrated in Fig.~\ref{fig:1DbinaryMERA}d, depend on the measurement outcomes $\vr$ in the importance sampling. If we want the optimization to reach energy-accuracy $\epsilon$, we need
\begin{equation}\label{eq:VMC-NoSamples}
	\O(1/\epsilon^2)\stackrel{\eqref{eq:eps-chi}}{=}\O(\chi^{2\beta})
\end{equation}
samples to get a sufficiently accurate gradient estimate. The contraction sequences for the gradient tensor networks \eqref{eq:VMC-grad} can again be optimized by the breadth-first constructive search (Fig.~\ref{alg:breadthFirst}). When employing EMB \eqref{eq:EMB}, we include the basis states $|r\ket$ as separate elements of the tensor network. For the CMB, we simply remove the measured indices from the corresponding tensors.

Compared to the CMB, the need to evaluate reduced density matrices in importance sampling with EMB causes additional costs. In Secs.~\ref{sec:fMERA-results}, \ref{sec:TMERA-results}, and \ref{sec:phaseDiagrams}, we analyze VMC using the CMB. Section~\ref{sec:EMB} discusses the cost scaling for VMC with EMB.

\subsection{Results}\label{sec:fMERA-results}
\begin{figure*}[p]
	\includegraphics[width=0.99\textwidth]{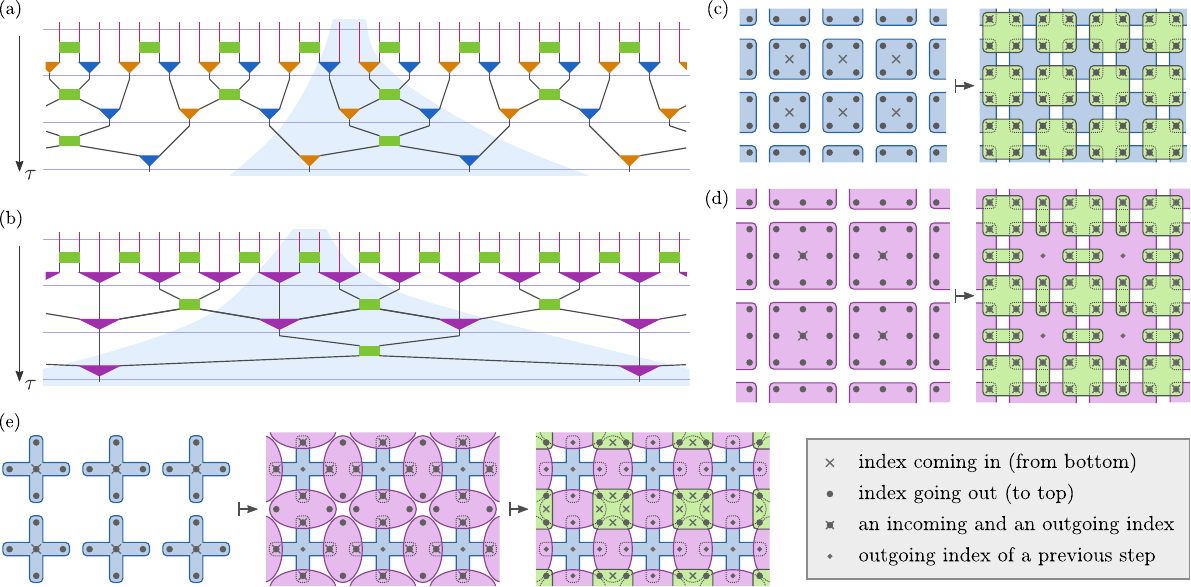}
	\caption{\label{fig:MERAs}\textbf{1D and 2D MERA tensor networks.} Panels (a) and (b) show three layers of the 1D modified binary MERA \cite{Evenbly2013} and 1D ternary MERA \cite{Evenbly2009-79}, respectively, where the preparation direction is upwards (decreasing $\tau$) and causal cones for local operators are indicated. Panels (c-e) each show one layer of a 2D MERA, where we progress in the preparation direction (decreasing $\tau$) from left to right. (c) For the shown 2D $2\times 2\mapsto 1$ MERA \cite{Cincio2008-100}, isometries map renormalized sites (crosses) into blocks of $2\times 2$ sites (dots) before disentanglers are applied to shifted $2\times 2$ site blocks. (d) For the shown 2D two-step $3\times 3\mapsto 1$ MERA \cite{Evenbly2009-79}, isometries map renormalized sites into blocks of $3\times 3$ sites before one applies $4$-site and two $2$-site disentanglers. (e) For the shown 2D three-step $3\times 3\mapsto 1$ MERA \cite{Evenbly2009-102}, isometries map renormalized sites into groups of five sites, applies isometries that map groups of two sites into four sites, and, finally, applies $4$-site disentanglers (also rearranging site locations).}
\end{figure*}
\begin{table*}[p]
	\setlength{\tabcolsep}{1.57ex}
	\renewcommand{\arraystretch}{1.45}
	\begin{tabular}{|l | c | c c | c |}
	\hline
	\multicolumn{1}{|c|}{         } & fMERA, exact $E$ gradients & \multicolumn{2}{c|}{fMERA, variational Monte Carlo (CMB)} & Comparison\\[-0.5em]
	\multicolumn{1}{|c|}{MERA type} & layer trans.\ \& energy env.   & importance sampling &  energy env. & $\beta$ bound\\
	\hline
	1D binary                & \dom{$\O(\chi^9)$}    & \jon{$\O(\chi^6/\epsilon^2)$}  & $\O(\chi^5/\epsilon^2)$  & $\beta<1.5$\\
	1D mod.\ binary, left    & \dom{$\O(\chi^7)$}    & \jon{$\O(\chi^5/\epsilon^2)$}  & $\O(\chi^4/\epsilon^2)$  & $\beta<1$\\[-0.5em]
	1D mod.\ binary, central &      $\O(\chi^6)$     &      $\O(\chi^4/\epsilon^2)$   & $\O(\chi^4/\epsilon^2)$  & \tqs\\[-0.5em]
	1D mod.\ binary, odd     &      $\O(\chi^6)$     &      $\O(\chi^4/\epsilon^2)$   & $\O(\chi^3/\epsilon^2)$  & \tqs\\
	1D ternary, left         & \dom{$\O(\chi^8)$}    & \jon{$\O(\chi^5/\epsilon^2)$}  & $\O(\chi^4/\epsilon^2)$  & $\beta<1.5$\\[-0.5em]
	1D ternary, central      &      $\O(\chi^6)$     & \jon{$\O(\chi^5/\epsilon^2)$}  & $\O(\chi^4/\epsilon^2)$  & \tqs\\
	2D $2\times 2\mapsto 1$  & \dom{$\O(\chi^{26})$} & \jon{$\O(\chi^{16}/\epsilon^2)$}  & $\O(\chi^{14}/\epsilon^2)$  & $\beta<5$\\
	2D two-step $3\times 3\mapsto 1$, TL & \dom{$\O(\chi^{16})$}   & \jon{$\O(\chi^{15}/\epsilon^2)$}  & $\O(\chi^{10}/\epsilon^2)$  & $\beta<0.5$\\[-0.5em]
	2D two-step $3\times 3\mapsto 1$, TC &      $\O(\chi^{15})$    &      $\O(\chi^{14}/\epsilon^2)$   & $\O(\chi^{10}/\epsilon^2)$  & \tqs\\[-0.5em]
	2D two-step $3\times 3\mapsto 1$, MC &      $\O(\chi^{12})$    &      $\O(\chi^{13}/\epsilon^2)$   & $\O(\chi^{10}/\epsilon^2)$  & \tqs\\
	2D three-step $3\times 3\mapsto 1$, TL & \dom{$\O(\chi^{16})$} & \jon{$\O(\chi^{11}/\epsilon^2)$}  & $\O(\chi^{8}/\epsilon^2)$  & $\beta<2.5$\\[-0.5em]
	2D three-step $3\times 3\mapsto 1$, TC &      $\O(\chi^{15})$  &      $\O(\chi^{10}/\epsilon^2)$   & $\O(\chi^{8}/\epsilon^2)$  & \tqs\\[-0.5em]
	2D three-step $3\times 3\mapsto 1$, MC &      $\O(\chi^{14})$  &      $\O(\chi^{10}/\epsilon^2)$   & $\O(\chi^{8}/\epsilon^2)$  & \tqs\\
	\hline
	\end{tabular}
	\caption{\label{tab:fMERA}\textbf{Contraction costs for full MERA.} For different types of fMERA, i.e., those with non-Trotterized tensors, we list the optimal scaling of contraction costs. The most costly steps in the EEG and VMC approaches are indicated in red and blue, respectively. When a MERA involves different types of layer transitions, we list the costs for all inequivalent transition types. For VMC, we have multiplied with the required number of samples  \eqref{eq:VMC-NoSamples} to obtain the total cost per iteration. With the energy-density error scaling as $\epsilon\sim \chi^{-\beta}$, the last column gives the maximum $\beta$ for which VMC can be advantageous over EEG-based optimization. Contraction sequences with the optimal cost scaling are provided in Appendices~\ref{appx:1DbinaryMERA}-\ref{appx:2D3s3x3to1MERA}. While we consider VMC employing the CMB here, VMC with EMB is discussed in Sec.~\ref{sec:EMB}.}
\end{table*}
Table~\ref{tab:fMERA} summarizes the scaling of fMERA contraction costs for the EEG and VMC energy-optimization algorithms with corresponding optimal contraction sequences given in Appendices~\ref{appx:1DbinaryMERA}-\ref{appx:2D3s3x3to1MERA}. We consider the following six types of 1D and 2D MERA displayed in Figs.~\ref{fig:MERA-Trotter} and \ref{fig:MERAs}:
\begin{itemize}
 \item The 1D binary MERA with branching ratio $b=2$ \cite{Vidal-2005-12} has causal-cone width $A=3$ and two equivalent layer transitions (left and right-moving); see Appx.~\ref{appx:1DbinaryMERA}).
 \item The 1D modified binary MERA with branching ratio $b=2$ \cite{Evenbly2013} has causal-cone width $A=2$ and four layer transitions -- the equivalent left and right-moving ones (mapping from even to odd bonds), a central one (from even to even bonds), and an odd transition (from odd to even bonds); see Appx.~\ref{appx:1DbinaryModMERA}.
 \item The 1D ternary MERA with branching ratio $b=3$ \cite{Evenbly2009-79} has causal-cone width $A=2$ and three layer transitions -- the equivalent left and right-moving ones and a central transition; see Appx.~\ref{appx:1DternaryMERA}.
 \item The 2D quaternary MERA mapping blocks of $2\times 2$ sites into one renormalized site ($2\times 2\mapsto 1$) with branching ratio $b=4$ \cite{Cincio2008-100} has causal-cone width $A=3\times 3$ and the four equivalent layer transitions $\E_\TL,\E_\TR,\E_\BL$, and $\E_\BR$; see Appx.~\ref{appx:2D2x2to1MERA}.
 \item The 2D nonary MERA mapping blocks of $3\times 3$ sites in two steps into one renormalized site ($3\times 3\mapsto 1$) with branching ratio $b=9$  \cite{Evenbly2009-79} has causal-cone width $A=2\times 2$ and nine layer transitions which fall into the three inequivalent classes $\{\E_\TL,\E_\TR,\E_\BL,\E_\BR\}$, $\{\E_\TC,\E_\ML,\E_\MR,\E_\BC\}$, and $\{\E_\MC\}$; see Appx.~\ref{appx:2D2s3x3to1MERA}.
 \item The 2D nonary MERA mapping blocks of $3\times 3$ sites in three steps into one renormalized site ($3\times 3\mapsto 1$) with branching ratio $b=9$ \cite{Evenbly2009-102} has causal-cone width $A=2\times 2$ and nine layer transitions which fall into the three inequivalent classes $\{\E_\TL,\E_\TR,\E_\BL,\E_\BR\}$, $\{\E_\TC,\E_\ML,\E_\MR,\E_\BC\}$, and $\{\E_\MC\}$; see Appx.~\ref{appx:2D3s3x3to1MERA}.
\end{itemize}

The tensor contractions for layer transitions \eqref{eq:prop-EEG} and all environment tensors $\partial_{U_k}\bra\Psi|\hh_i|\Psi\ket$ in the EEG approach as described in the introduction (Sec.~\ref{sec:intro}) and Ref.~\cite{Evenbly2009-79} turn out to always have the same cost scaling:
\begin{lemma}\label{lemma4}
  We are given a closed tensor network, i.e., a tensor network without uncontracted indices. Then,
  (a) the single-tensor environment tensors obtained by removing one of the elementary tensors from the network all have the same minimal contraction cost $\O(\chi^\alpha)$. Furthermore, 
  (b) given an optimal contraction sequence for one environment tensor, we can deduce optimal contraction sequences for all environments, and
  (c) caching tensors of intermediate contraction steps, the environments of all tensors can be computed at cost $\O(\chi^\alpha)$.
\end{lemma}
Based on Ref.~\cite{Evenbly2014-89}, Appx.~\ref{appx:lemma4} provides a short proof, which is based on cost-conserving transformations among contraction trees. The tensor networks for layer transitions \eqref{eq:prop-EEG} and environment tensors can all be obtained from the closed tensor network for expectation values $\bra\Psi|\hh_i|\Psi\ket=\Tr\big[\hh_i^{\tau-i}\E_{\tau,i}(\dm_i^\tau)\big]$ by removing a single tensor; either $\hh_i^{\tau-i}$ or one isometry or disentangler from layer $\tau$. Hence, they have the same cost scaling.

Due to Lemma~\ref{lemma4} there is only one column in Table~\ref{tab:fMERA} giving the EEG contraction costs for layer transitions and environment tensors. For MERA with inequivalent layer transitions, we list the costs for each class. Contraction sequences with the optimal cost scaling are provided in Appendices~\ref{appx:1DbinaryMERA}-\ref{appx:2D3s3x3to1MERA}. We provide an optimal contraction sequence for one of the environment tensors and all others can be deduced through cost-conserving transformations of the corresponding contraction tree; see Appx.~\ref{appx:lemma4} and Ref.~\cite{Evenbly2014-89}. 
For the 1D fMERA, the optimal EEG contraction costs scale as
\begin{equation}
	\O(\chi^9),\quad
	\O(\chi^7),\quad\text{and}\quad
	\O(\chi^8)
\end{equation}
for the binary, modified binary, and ternary 1D MERA, respectively \cite{Evenbly2013}.
For the 2D fMERA, the costs scale as
\begin{equation}
	\O(\chi^{26}),\quad
	\O(\chi^{16}),\quad\text{and}\quad
	\O(\chi^{16})
\end{equation}
for the quaternary \cite{Pfeifer2014-90}, two-step nonary \cite{Evenbly2009-79}, and three-step nonary 2D fMERA \cite{Evenbly2009-102}, respectively.

In comparison, the dominant contraction costs for the fMERA VMC algorithm with CMB, always incurred in the importance sampling step, are found to scale as 
\begin{equation}
	\O(\chi^6/\epsilon^2),\quad \,\
	\O(\chi^5/\epsilon^2),\quad \ \text{and}\quad
	\O(\chi^5/\epsilon^2)
\end{equation}
in 1D and as 
\begin{equation}
	\O(\chi^{16}/\epsilon^2),\quad
	\O(\chi^{15}/\epsilon^2),\quad\text{and}\quad
	\O(\chi^{11}/\epsilon^2)
\end{equation}
for 2D. As discussed in Sec.~\ref{sec:fMERA-VMC}, we need $\O(1/\epsilon^2)$ samples for a sufficiently precise VMC energy (gradient) evaluation which, for critical systems, means $\O(\chi^{2\beta})$ samples according to Eq.~\eqref{eq:eps-chi}.

To decide whether the EEG approach or the VMC approach is more efficient, we need to take into account the dominant contraction costs and the \emph{tensor update cost}. For the latter, consider a tensor $U_k$ mapping from $\nu$ input sites to $\mu\geq\nu$ output sites. The tensor and its energy gradient (environment tensor) $d_k:=\partial_{U_k}\bra\Psi|\hh_i|\Psi\ket$ are $\chi^\mu\times\chi^\nu$ matrices and a naive gradient-descent update $U_k\mapsto U_k-\gamma d_k$ would have a cost of $\O(\chi^{\mu+\nu})$. However, all tensors need to stay isometric, i.e., obey $U_k^\dag U_k=\id_{\chi^\nu}$. In practice, this is done in two ways:
\begin{enumerate}
 \item[(a)] In Riemannian gradient descent or quasi-Newton methods as described in Refs.~\cite{Hauru2021-10,Luchnikov2021-23,Miao2021_08}, one projects the gradient $d_k$ onto the tangent space of the relevant Stiefel manifold, and employs retractions and/or parallel transport on that manifold.
 \item[(b)] In the more traditional fixed-point iteration as described in Ref.~\cite{Evenbly2009-79}, one treats the energy functional as a linear function of independent tensors $U_k$ and $U_k^\dag$, and employs a singular value decompositions of the environment tensor $d_k$ to update $U_k$.
\end{enumerate}
In both cases, the tensor update cost scales as
\begin{equation}\label{eq:tensor-update}
	\O(\chi^{\mu+2\nu}),
\end{equation}
which is $\O(\chi^6)$ for the 1D MERA and $\O(\chi^{12})$ for the 2D MERA. So, if $\beta<1/2$, it is the dominant cost for fMERA VMC with the 1D modified binary, 1D ternary, and 2D three-step $3\times 3\mapsto 1$ MERA, and it is subleading in all other cases.

The last column in Table~\ref{tab:fMERA} states the resulting model-dependent bounds for $\beta$ below which VMC is expected to be more efficient than the EEG approach. Of course, there are coefficients to the leading cost terms and subleading corrections that we ignored. They can make a difference at small bond dimensions $\chi$. For 1D modified binary MERA, we have the bound $\beta<1$. However, for various critical spin chains, we found $\beta\gtrsim 1.2$ as summarized in Table~\ref{tab:bp}. The $\beta$ bounds in Table~\ref{tab:fMERA} suggest a particularly big separation between VMC and EEG for the 2D quaternary and 2D three-step nonary MERA. Unfortunately, 2D MERA EEG simulations are quite expensive such that (numerical) data for $\beta$ in 2D systems is currently not yet available. It might be possible to estimate these energy-accuracy exponents for critical 2D systems using VMC.

\section{TMERA contraction costs}\label{sec:TMERA}
\subsection{Trotterized MERA}
To implement MERA efficiently on quantum computers \cite{Kim2017_11,Miao2021_08,Haghshenas2022-12,Miao2023_03,Barthel2023_03,Miao2024-109,Haghshenas2023_05,Job2024_04,Miao2024_12} one can impose a substructure on the MERA tensors, associating $q$ qubits to each renormalized site such that the bond dimension is
\begin{equation}
	\chi=2^q 
\end{equation}
and realizing each (isometric or unitary) tensor as a circuit of quantum gates as described in the introduction~\ref{sec:intro}. A simple choice are brickwall circuits which comprise $\propto t$ layers of (complete) two-qubit-gate coverings and where the exact number of gate layers is a small-integer multiple of $t$, e.g., 2 for 1D TMERA and 4 for 2D TMERA with brickwall circuits and a square-lattice qubit arrangement. See Fig.~\ref{fig:MERA-Trotter}c. We call $t$ the number of Trotter steps. For simplicity, we assume the same $t$ for all tensors in a TMERA.
For critical systems, the optimal number of Trotter steps increases polynomially in the bond dimension \cite{Miao2021_08,Miao2023_03} according to $t\sim\chi^p$ [Eq.~\eqref{eq:t-chi}]. An fMERA tensor that maps from $\nu$ input sites to $\mu$ output sites, each associated with a $\chi$-dimensional bond vector space, can always be implemented as a circuit of $\O(\chi^{\nu+\mu})$ single-qubit and CNOT gates \cite{Iten2016-93}. This means that
\begin{equation}\label{eq:pMax}
	p\leq \nu+\mu
\end{equation}
is sufficient for the TMERA tensors. This upper bound on the exponent $p$ in Eq.~\eqref{eq:t-chi} is stated for various TMERA in the $p_\text{max}$ columns of Tables~\ref{tab:TMERA} and \ref{tab:EMB}.

\subsection{Pairwise contractions with Trotterized tensors}\label{sec:TMERA-contract}
In several cases, networks with Trotterized tensors can be contracted more efficiently than networks of unconstrained tensors, also on classical computers.

\begin{figure}[t]
	\includegraphics[width=1\columnwidth]{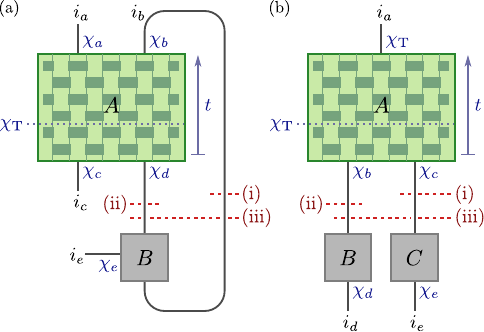}
	\caption{\label{fig:TMERA-contract}\textbf{Contractions with Trotterized tensors.}
	Contractions with Trotterized tensors are generally more efficient than contractions with unconstrained tensors. (a) Depending on the bond dimensions, one of the three indicated sequences for the pairwise contraction of a Trotterized tensor $A$ and a generic tensor $B$ will be optimal as discussed in Sec.~\ref{sec:TMERA-contract}. (b) With Trotterization, we also need to consider outer products of tensors connected to the same side of a Trotterized tensor as discussed in Sec.~\ref{sec:TMERA-costOpt}.}
\end{figure}

\Emph{A Trotterized and a generic tensor}
Consider the tensor network in Fig.~\ref{fig:TMERA-contract}a, which represents the most general pairwise contraction
\begin{equation}\label{eq:contract-A-B}
	\sum_{i_b=1}^{\chi_b}\sum_{i_d=1}^{\chi_d}A_{i_a,i_b,i_c,i_d}B_{i_b,i_d,i_e}
\end{equation}
of a Trotterized tensor $A$ with a generic tensor $B$, where $i_a\in[1,\chi_a]$, $i_c\in[1,\chi_c]$, and $i_e\in[1,\chi_e]$. Let the circuit of $A$ comprise $t$ Trotter steps and let $\chi_\tT$ be the maximum dimension of the vector space that the gate layers act on; due to possible projections on the input and/or output sides of $A$ (e.g., when $A$ is a partial isometry or when a projective measurement is applied in the VMC approach), we have
\begin{equation}\label{eq:TrotterDim}
	\chi_\tT\geq \max(\chi_a\chi_b,\chi_c\chi_d).
\end{equation}

Without Trotterization, the contraction cost would simply be
\begin{equation}\label{eq:TrotterContract-0}
	\O(\chi_a\chi_b\chi_c\chi_d\chi_e).
\end{equation}

With the Trotter structure of $A$ there are three possible contraction sequences as indicated in Fig.~\ref{fig:TMERA-contract}a:
\begin{itemize}
\item[(i)] 
  We initially keep $i_b$ uncontracted and contract $A$'s Trotter circuit bottom-up, layer by layer. Each contraction of one two-qubit gate at the current top end of the partially contracted circuit costs $\O(\chi_b\chi_c\chi_e\cdot\chi_\tT)$ operations.
  In the second step, we sum over $i_b$. The total cost is
  \begin{equation}\label{eq:TrotterContract-1}
  	\O(t \chi_b\chi_c\chi_e\chi_\tT+\chi_a\chi_b\chi_c\chi_e).
  \end{equation}
\item[(ii)] 
  We initially keep $i_d$ uncontracted and contract $A$'s Trotter circuit top-down, layer by layer. Each contraction of one two-qubit gate at the current bottom end of the partially contracted circuit costs $\O(\chi_a\chi_d\chi_e\cdot\chi_\tT)$ operations.
  In the second step, we sum over $i_d$. The total cost is
  \begin{equation}\label{eq:TrotterContract-2}
  	\O(t \chi_a\chi_d\chi_e\chi_\tT+\chi_a\chi_c\chi_d\chi_e).
  \end{equation}
\item[(iii)] 
  We can initially keep both $i_b$ and $i_d$ uncontracted, contract $A$'s Trotter circuit, and, finally, execute the sums over $i_b$ and $i_d$ simultaneously. In the latter step, $A$ behaves like an unconstrained tensor such that the associated contractions cost is simply \eqref{eq:TrotterContract-0}. The total cost is
  \begin{equation}\label{eq:TrotterContract-3}
  	\O(t \min(\chi_a\chi_b,\chi_c\chi_d)\chi_\tT+\chi_a\chi_b\chi_c\chi_d\chi_e).
  \end{equation}
\end{itemize}
\begin{table*}[p]
	\setlength{\tabcolsep}{1.5ex}
	\begin{tabular}{l c c c c}
	\hline
	Model & central charge $c$  & energy exponent\ $\beta$ & Trotter exponent\ $p$ & In Fig.~\ref{fig:algCmp}b\\[0.5em]
	\hline
	Spin-1/2 XX chain		& 1	& 3.68 & 1.66		& green\\
	Spin-1/2 XXX chain		& 1	& 2.8  & 2.1		& red\\
	Spin-1 BLBQ chain, TB point	& 3/2	& 1.75 & 1.78		& blue\\
	Spin-1 BLBQ chain, ULS point	& 2	& 1.19\dots1.74 & 1.89	& purple\\
	Spin-3/2 BLBQBC chain, TB point	& 9/5	& 1.48 & 1.54		& brown\\
	Spin-3/2 XXX chain		& 1	& 1.75 & 1.31		& orange\\
	\hline
	\end{tabular}
	\caption{\label{tab:bp}\textbf{MERA scaling exponents for critical spin chains.} For critical models, the energy accuracy $\epsilon$ and optimal number of Trotter steps $t$ follow power laws in terms of the bond dimension $\chi$ as stated in Eqs.~\eqref{eq:eps-chi} and \eqref{eq:t-chi}. The table lists scaling exponents $\beta$ and $p$ for the 1D modified binary MERA as determined by numerical simulation in Ref.~\cite{Miao2023_03}, where the Trotter circuits have brickwall structure. The considered models are the isotropic spin-1/2 Heisenberg antiferromagnet \cite{Bethe1931,Hulthen1938,Cloizeaux1966-7,Johnson1972-6,Mikeska2004}, the bilinear-biquadratic (BLBQ) spin-1 chain \cite{Uimin1970-12,Lai1974-15,Sutherland1975-12,Takhtajan1982-87,Babujian1982-90,Babujian1983-215,Laeuchli2006-74,Binder2020-102} at the integrable Takhtajan-Babujian \cite{Takhtajan1982-87,Babujian1982-90,Babujian1983-215} and Uimin-Lai-Sutherland points \cite{Uimin1970-12,Lai1974-15,Sutherland1975-12}, the bilinear-biquadratic-bicubic (BLBQBC) spin-3/2 chain at the Takhtajan-Babujian point \cite{Takhtajan1982-87,Babujian1982-90,Babujian1983-215,Alcaraz1988-21}, and the isotropic spin-$3/2$ Heisenberg antiferromagnet \cite{Haldane1983-93,Haldane1983-50,Schulz1986-33,Schulz1986-34,Affleck1987-36,Alcaraz1992-46,Hallberg1996-76}.}
\end{table*}
\begin{table*}[p]
	\setlength{\tabcolsep}{1.4ex}
	\renewcommand{\arraystretch}{2.4}
	\begin{tabular}{|l | c | l | l l |}
	\hline
	\multicolumn{1}{|c|}{         } &                & \multicolumn{1}{c|}{TMERA, EEG} & \multicolumn{2}{c|}{TMERA, variational Monte Carlo (CMB)}\\[-1.5em]
	\multicolumn{1}{|c|}{MERA type} & $p_\text{max}$ & \multicolumn{1}{|c|}{layer trans.\ \& energy env.} & importance sampling &  energy environment\\
	\hline
	1D binary                & 4   & \dom{$\O\Big(\chi^{8\lin{0}{1}9}\Big)$}            & \jon{$\O(\chi^{4+p}/\epsilon^2)$}  & $\O\Big(\chi^{\jon{4\lin{0}{1}}5\lin{2}{4}7}/\epsilon^2\Big)$\\
	1D mod.\ binary, left    & 4   & \dom{$\O\Big(\chi^{6\lin{0}{1}7\lin{3}{4}8}\Big)$} & \jon{$\O(\chi^{3+p}/\epsilon^2)$}  & \jon{$\O(\chi^{3+p}/\epsilon^2)$}\\[-0.6em]
	1D mod.\ binary, central & \tq &      $\O\Big(\chi^{6\lin{2}{4}8}\Big)$             & \jon{$\O(\chi^{3+p}/\epsilon^2)$}  & \jon{$\O(\chi^{3+p}/\epsilon^2)$}\\[-0.6em]
	1D mod.\ binary, odd     & \tq &      $\O\Big(\chi^{6\lin{3}{4}7}\Big)$             & \jon{$\O(\chi^{3+p}/\epsilon^2)$}  & \jon{$\O(\chi^{3+p}/\epsilon^2)$}\\
	1D ternary, left         & 4   & \dom{$\O(\chi^{8})$}                     & \jon{$\O(\chi^{4+p}/\epsilon^2)$}  & \jon{$\O(\chi^{4+p}/\epsilon^2)$}\\[-0.6em]
	1D ternary, central      & \tq &      $\O\Big(\chi^{6\lin{2}{4}8}\Big)$   & \jon{$\O(\chi^{4+p}/\epsilon^2)$}  & \jon{$\O(\chi^{4+p}/\epsilon^2)$}\\
	2D $2\times 2\mapsto 1$  & 8   & \dom{$\O(\chi^{26})$}  & \jon{$\O\Big(\chi^{14\lin{0}{2}16}/\epsilon^2\Big)$}  & $\O\Big(\chi^{13\lin{1}{2}14\lin{6}{8}16}/\epsilon^2\Big)$\\
	2D two-step $3\times 3\mapsto 1$, TL & 10  & \dom{$\O\Big(\chi^{16\lin{6}{10}20}\Big)$}            & \jon{$\O\Big(\chi^{14\lin{0}{1}15\lin{5}{10}20}/\epsilon^2\Big)$} & $\O(\chi^{10+p}/\epsilon^2)$\\[-0.6em]
	2D two-step $3\times 3\mapsto 1$, TC & \tq &      $\O\Big(\chi^{14\lin{2}{3}15\lin{5}{10}20}\Big)$ &     {$\O\Big(\chi^{13\lin{0}{1}14\lin{4}{10}20}/\epsilon^2\Big)$} & $\O(\chi^{10+p}/\epsilon^2)$\\[-0.6em]
	2D two-step $3\times 3\mapsto 1$, MC & \tq &      $\O\Big(\chi^{12\lin{2}{10}20}\Big)$             &     {$\O\Big(\chi^{12\lin{0}{1}13\lin{3}{10}20}/\epsilon^2\Big)$} & $\O(\chi^{10+p}/\epsilon^2)$\\
	2D three-step $3\times 3\mapsto 1$, TL & 8   & \dom{$\O(\chi^{16})$}                                & \jon{$\O\Big(\chi^{9\lin{0}{2}11\lin{4}{8}15}/\epsilon^2\Big)$} & $\O\Big(\chi^{8\lin{2}{8}14}/\epsilon^2\Big)$\\[-0.6em]
	2D three-step $3\times 3\mapsto 1$, TC & \tq &      $\O\Big(\chi^{14\lin{0}{1}15\lin{7}{8}16}\Big)$ &     {$\O\Big(\chi^{9\lin{0}{1}10\lin{4}{8}14}/\epsilon^2\Big)$} & $\O\Big(\chi^{7\lin{0}{1}8\lin{2}{8}14}/\epsilon^2\Big)$\\[-0.6em]
	2D three-step $3\times 3\mapsto 1$, MC & \tq &      $\O\Big(\chi^{14\lin{6}{8}16}\Big)$             &     {$\O\Big(\chi^{9\lin{0}{1}10\lin{4}{8}14}/\epsilon^2\Big)$} & $\O\Big(\chi^{7\lin{1}{8}14}/\epsilon^2\Big)$\\
	\hline
	\end{tabular}
	\caption{\label{tab:TMERA}\textbf{Contraction costs for Trotterized MERA.} For different types of Trotterized MERA, i.e., those with tensors consisting of circuits with $t\sim \chi^p$ layers of quantum gates, we list the optimal scaling of contraction costs. The most costly contraction steps in the EEG optimization and VMC approaches are indicated in red and blue, respectively. As a function of $p$, the exponents have constant and linearly increasing intervals. In the notation above, ``$\alpha_1\lin{p_1}{p_2}\alpha_2$'' means that the exponent has value $\alpha_1$ until $p=p_1$ and changes linearly on the interval $(p_1,p_2)$ reaching the value $\alpha_2$ at $p=p_2$ (and staying there for $p>p_2$). For VMC, we have multiplied with the required number of samples $\O(1/\epsilon^2)$ to obtain the total cost per iteration, where $\epsilon\sim \chi^{-\beta}$. While we consider VMC employing the CMB here, VMC with EMB is discussed in Sec.~\ref{sec:EMB}.}
\end{table*}

\Emph{Two Trotterized tensors}
If $A$ and $B$ in Eq.~\eqref{eq:contract-A-B} are both Trotterized tensors, the optimal contraction will consist in, first, contracting the Trotter circuit of one of the tensors (making it a generic tensor like $B$ above) and, second, applying one of the three strategies discussed in the previous paragraph. This is due to the fact that, (a) for critical systems, the Trotter circuits are deep in the sense that the number $t$ of Trotter steps (gate layers) scales polynomially in the input and out dimensions according to Eq.~\eqref{eq:t-chi} and (b), due to Eq.~\eqref{eq:TrotterDim}. For optimal contraction sequences this implies that, once, we have started contracting the Trotter circuits of one tensor, we always can/should complete this circuit contraction before further steps.

\Emph{Beyond pairwise contractions}
For the same reasons and Lemma~\ref{lemma2}, it should never be necessary to go beyond pairwise contractions, also with Trotterized tensors. At least for the networks considered in this work, we did not find a situation where this would be beneficial.

\subsection{Contraction-cost optimization}\label{sec:TMERA-costOpt}
To determine optimized contraction sequences for networks with Trotterized tensors, we can again employ the breadth-first constructive search (Fig.~\ref{alg:breadthFirst}). In accordance with Sec.~\ref{sec:TMERA-contract}, we replace in the procedure \textproc{costExponent}{$(T,T')$}, Eq.~\eqref{eq:TrotterContract-0} by the minimum of Eqs.~\eqref{eq:TrotterContract-1}-\eqref{eq:TrotterContract-3} when contracting a Trotterized and a generic tensor. When contracting two Trotterized tensors $A$ and $B$, we add the Trotter-circuit contraction cost for $B$ to the minimum of Eqs.~\eqref{eq:TrotterContract-1}-\eqref{eq:TrotterContract-3} and minimize over the two orderings of the two tensors (assignments of labels ``$A$'' and ``$B$'').

There is one more complication in comparison to the case of networks with unconstrained tensors (Sec.~\ref{sec:fMERA}): For two tensors $B$ and $C$ which are both contracted to the input or the output side of a Trotterized tensor $A$, it may be advantageous to first form their outer product before the contraction with $A$. Specifically, consider the contraction
\begin{equation}
	\sum_{i_b=1}^{\chi_b}\sum_{i_c=1}^{\chi_c}A_{i_a,i_b,i_c}B_{i_b,i_d}C_{i_c,i_e}
\end{equation}
as illustrated in Fig.~\ref{fig:TMERA-contract}b, where $i_a\in[1,\chi_\tT]$, $i_d\in[1,\chi_d]$, and $i_e\in[1,\chi_e]$.
The three possible contraction sequences and associated costs are
\begin{subequations}
\begin{align}
	&(A\,B)\,C:\quad\O(t\chi_c\chi_d \chi_\tT + \chi_c\chi_d\chi_e\chi_\tT),\\
	&(A\,C)\,B:\quad\O(t\chi_b\chi_e \chi_\tT + \chi_b\chi_d\chi_e\chi_\tT),\\
	&A\,(B\,C):\quad\O(\chi_b\chi_c\chi_d\chi_e + t \chi_d\chi_e \chi_\tT).
\end{align}
\end{subequations}
The third option with the outer product $B\,C$ can have the lowest cost. This is, for example, the case, when $\chi_d=\chi_e=\chi$, $\chi_b=\chi_c=\chi^2$, $\chi_\tT=\chi^4$, and $t\propto \chi\dots\chi^{8}$.
Hence, we also modify the procedure \textproc{contractible}{$(T,T')$} in the search algorithm, allowing for outer products of tensors if they are connected to the same side (input or output) of a Trotterized tensor.

\subsection{Results}\label{sec:TMERA-results}
Table~\ref{tab:TMERA} summarizes the scaling of TMERA contraction costs for the EEG and VMC energy-optimization algorithms, considering the same six types of 1D and 2D MERA as in Sec.~\ref{sec:fMERA-results}. In general, the optimal contraction sequences now depend on the exponent $p$ in the scaling relation \eqref{eq:t-chi} for the number of Trotter steps $t$. We specify optimal sequences for TMERA EEG layer transitions and TMERA VMC importance sampling in Appendices~\ref{appx:1DbinaryMERA}-\ref{appx:2D3s3x3to1MERA}.

For the 1D TMERA, the optimal EEG contraction costs scale as
\begin{equation}
	\O\big(\chi^{8\lin{0}{1}9}\big),\
	\O\big(\chi^{6\lin{0}{1}7\lin{3}{4}8}\big),\ \text{and}\ \
	\O\big(\chi^{8}\big)
\end{equation}
for the binary, modified binary, and ternary 1D TMERA, respectively. Here ``$\alpha_1\lin{p_1}{p_2}\alpha_2$'' means that the exponent has the value $\alpha_1$ for $p\leq p_1$, changes linearly on the interval $(p_1,p_2)$ reaching the value $\alpha_2$ at $p=p_2$, and stays there for $p>p_2$.
For the 2D TMERA, the optimal EEG contraction costs scale as
\begin{equation}
	\O\big(\chi^{26}\big),\quad
	\O\big(\chi^{16\lin{6}{10}20}\big),\quad \text{and}\quad
	\O\big(\chi^{16}\big)
\end{equation}
for the quaternary, two-step nonary, and three-step nonary 2D TMERA, respectively. In comparison to the results in Sec.~\ref{sec:fMERA-results}, this suggests that TMERA EEG on classical computers can only outperform fMERA EEG for the 1D binary and modified binary MERA when $p$ is below one. In view of Table~\ref{tab:bp}, $p<1$ seems untypical.
\begin{figure*}[t]
        \setlength{\tabcolsep}{2ex}
	\begin{tabular}{l l l}
	{\footnotesize (a) \ 1D binary MERA}&
	{\footnotesize (b) \ 1D modified binary MERA}&
	{\footnotesize (c) \ 1D ternary MERA}\\[0.2em]
	\includegraphics[width=0.3\textwidth]{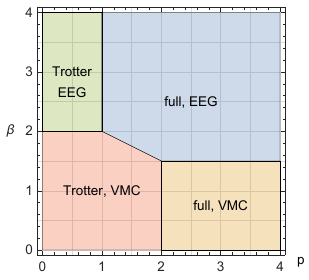}&
	\includegraphics[width=0.3\textwidth]{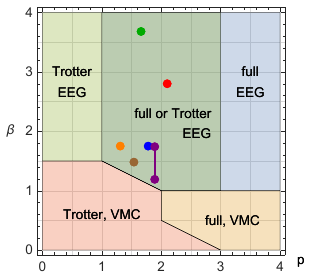}&
	\includegraphics[width=0.3\textwidth]{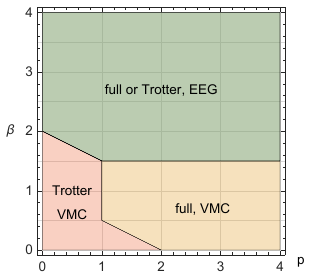}\\[0.6em]
	{\footnotesize (d) \ 2D $2\times 2\mapsto 1$ MERA}&
	{\footnotesize (e) \ 2D two-step   $3\times 3\mapsto 1$ MERA}&
	{\footnotesize (f) \ 2D three-step $3\times 3\mapsto 1$ MERA}\\[0.2em]
	\includegraphics[width=0.3\textwidth]{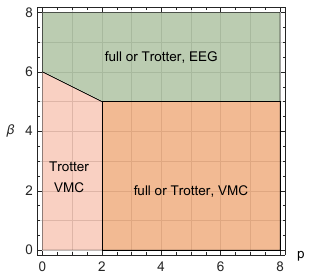}&
	\includegraphics[width=0.3\textwidth]{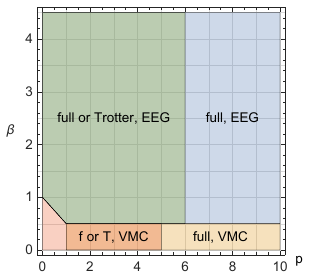}&
	\includegraphics[width=0.3\textwidth]{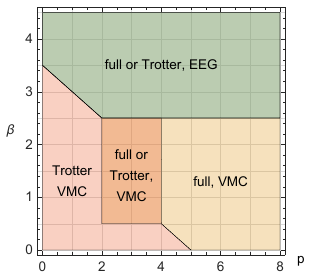}\\
	\end{tabular}
	\caption{\label{fig:algCmp}\textbf{Algorithmic phase diagrams.} For critical models, the energy accuracy $\epsilon$ and optimal number of Trotter steps $t$ follow power laws in terms of the bond dimension $\chi$ as stated in Eqs.~\eqref{eq:eps-chi} and \eqref{eq:t-chi}. In dependence of the corresponding MERA and model-specific scaling exponents $\beta$ and $p$, the results for the costs of optimal contraction sequences in Tables~\ref{tab:fMERA} and \ref{tab:TMERA} lead to the shown algorithmic phase diagrams. They specify which of the four algorithms in question (fMERA EEG, fMERA VMC, TMERA EEG, and TMERA VMC) has the smallest scaling exponent for the computation costs per optimization step. The colored points in panel (b) indicate the scaling exponents of different critical spin models as specified in Table~\ref{tab:bp}. The upper end of the shown range $0\leq p\leq p_{\max}$ is determined by Eq.~\eqref{eq:pMax}.}
\end{figure*}

The dominant contraction costs for the TMERA VMC algorithm always occur in the importance-sampling step and are found to scale as 
\begin{equation}\label{eq:TMERA-results-1D}
	\O(\chi^{4+p}/\epsilon^2),\ 
	\O(\chi^{3+p}/\epsilon^2),\ \ \text{and}\ \
	\O(\chi^{4+p}/\epsilon^2)
\end{equation}
in 1D and as
\begin{gather}\nonumber
	\O\Big(\chi^{14\lin{0}{2}16}/\epsilon^2\Big),\quad
	\O\Big(\chi^{14\lin{0}{1}15\lin{5}{10}20}/\epsilon^2\Big),\\
	\text{and}\quad
	\O\Big(\chi^{9\lin{0}{2}11\lin{4}{8}15}/\epsilon^2\Big)
	\label{eq:TMERA-results-2D}
\end{gather}
for 2D. Recall that, according to Eq.~\eqref{eq:eps-chi}, we need $\O(1/\epsilon^2)=\O(\chi^{2\beta})$ samples for the VMC energy (gradient) evaluation. This factor has been multiplied with the single-sample contraction costs in Eqs.~\eqref{eq:TMERA-results-1D} and \eqref{eq:TMERA-results-2D}.

When comparing the different methods, we also need to take into account the tensor-update costs as discussed in Sec.~\ref{sec:fMERA-results}. Luckily, they are always subleading for the TMERA: When applying tensor updates using the Riemannian approach or fixed-point iteration, the cost is simply proportional to the number of two-qubit gates in the Trotterized tensors. Up to logarithmic corrections, it is $\O(t/\epsilon^2)=\O(\chi^p/\epsilon^2)$.

\section{Algorithmic phase diagrams}\label{sec:phaseDiagrams}
Integrating the results from Tables~\ref{tab:fMERA} and \ref{tab:TMERA}, Fig.~\ref{fig:algCmp} provides algorithmic phase diagrams, which are the central result of this work. As a function of the scaling exponents $\beta$ and $p$ for the energy accuracy \eqref{eq:eps-chi} and number of Trotter steps \eqref{eq:t-chi}, the diagrams show which of the four algorithms (fMERA EEG, fMERA VMC, TMERA EEG, and TMERA VMC) will have the lowest computation costs per iteration in the energy optimization. Recall that, as we ignored coefficients of the leading cost terms and subleading terms, these bounds apply for large bond dimensions $\chi$. Generally, the fMERA EEG approach is most efficient when both $\beta$ and $p$ are above certain thresholds. The TMERA VMC approach is best when both exponents are small. TMERA EEG and fMERA VMC can prevail when $\beta$ or $p$, respectively, is large and the other of the two exponents is below a certain threshold.

For 1D modified binary MERA, we have, for example, the following boundaries: TMERA EEG is more efficient than fMERA EEG for
\begin{equation*}
	6\lin{0}{1}7\lin{3}{4}8 < 7\quad \Leftrightarrow\quad p< 1,
\end{equation*}
both have the same scaling for $1\leq p\leq 3$, and fMERA EEG is better for $p\geq 3$. fMERA VMC is more efficient than fMERA EEG for
\begin{equation*}
	\max\big(6,5+2\beta\big) < 7\quad \Leftrightarrow\quad \beta\leq 1.
\end{equation*}
TMERA VMC is more efficient than fMERA VMC for
\begin{equation*}
	3+p+2\beta < \max\big(6,5+2\beta\big),
\end{equation*}
which means $p\leq 2$ if $\beta\geq 1/2$.
While some of the studied critical spin chains have scaling exponents $p$ and $\beta$ close to these transition lines, all of them are located in the region where TMERA and fMERA EEG have the same cost scaling and outperform VMC; see Table~\ref{tab:bp} and Fig.~\ref{fig:algCmp}b. 

The phase diagrams suggest a particularly big separation between the VMC and EEG approaches for the 2D quaternary \cite{Cincio2008-100} and 2D three-step nonary MERA, with VMC being more efficient for $\beta\lesssim 5$ and $\beta\lesssim 2.5$, respectively. As an example, for $p\geq 2$ and $\beta=2$ with the 2D quaternary MERA, the EEG costs scale as $\O(\chi^{26})$ and the VMC costs as $\O(\chi^{20})$. For $2\leq p\leq 4$ and $\beta=1.5$ with the 2D three-step nonary MERA, the EEG costs scale as $\O(\chi^{16})$ and the VMC costs as $\O(\chi^{14})$.
\begin{table*}[t]
	\setlength{\tabcolsep}{1.4ex}
	\renewcommand{\arraystretch}{2.3}
	\begin{tabular}{|l | c c | c l l |}
	\hline
	\multicolumn{1}{|c|}{         } & \multicolumn{2}{c|}{fMERA, VMC (EMB-all)} & \multicolumn{3}{c|}{TMERA, VMC (EMB-all)}\\[-1.5em]
	\multicolumn{1}{|c|}{MERA type} & imp.\ sampling & energy env.   & $p_\text{max}$ & importance sampling & energy environment\\
	\hline
	1D binary                & \dom{$\O(\chi^6/\epsilon^2)$}  & $\O(\chi^5/\epsilon^2)$  & 4 & \jon{$\O\Big(\chi^{5\lin{1}{4}8}/\epsilon^2\Big)$}  & $\O\Big(\chi^{4\lin{0}{1}5\lin{2}{4}7}/\epsilon^2\Big)$\\
	1D mod.\ binary, left    & \dom{$\O(\chi^5/\epsilon^2)$}  & $\O(\chi^4/\epsilon^2)$  & 4   & \jon{$\O\Big(\chi^{4\lin{1}{4}7}/\epsilon^2\Big)$}  & {$\O(\chi^{3+p}/\epsilon^2)$}\\[-0.6em]
	1D mod.\ binary, central &      $\O(\chi^4/\epsilon^2)$   & $\O(\chi^4/\epsilon^2)$  & \tq & \jon{$\O\Big(\chi^{4\lin{1}{4}7}/\epsilon^2\Big)$}  & {$\O(\chi^{3+p}/\epsilon^2)$}\\[-0.6em]
	1D mod.\ binary, odd     &      $\O(\chi^4/\epsilon^2)$   & $\O(\chi^3/\epsilon^2)$  & \tq & \jon{$\O\Big(\chi^{4\lin{1}{4}7}/\epsilon^2\Big)$}  & {$\O(\chi^{3+p}/\epsilon^2)$}\\
	1D ternary, left         & \dom{$\O(\chi^6/\epsilon^2)$}  & $\O(\chi^4/\epsilon^2)$  & 4   & \jon{$\O\Big(\chi^{6\lin{2}{4}8}/\epsilon^2\Big)$}  & {$\O(\chi^{4+p}/\epsilon^2)$}\\[-0.6em]
	1D ternary, central      & \dom{$\O(\chi^6/\epsilon^2)$}  & $\O(\chi^4/\epsilon^2)$  & \tq & \jon{$\O\Big(\chi^{6\lin{2}{4}8}/\epsilon^2\Big)$}  & {$\O(\chi^{4+p}/\epsilon^2)$}\\
	2D $2\times 2\mapsto 1$  & \dom{$\O(\chi^{17}/\epsilon^2)$}  & $\O(\chi^{14}/\epsilon^2)$ & 8 & \jon{$\O(\chi^{17}/\epsilon^2)$}  & $\O\Big(\chi^{13\lin{1}{2}14\lin{6}{8}16}/\epsilon^2\Big)$\\
	2D two-step $3\times 3\mapsto 1$, TL & \dom{$\O(\chi^{19}/\epsilon^2)$}  & $\O(\chi^{10}/\epsilon^2)$  & 10  & \jon{$\O\Big(\chi^{19\lin{9}{10}20}/\epsilon^2\Big)$} & $\O(\chi^{10+p}/\epsilon^2)$\\[-0.6em]
	2D two-step $3\times 3\mapsto 1$, TC & {$\O(\chi^{17}/\epsilon^2)$}  & $\O(\chi^{10}/\epsilon^2)$  & \tq &      $\O\Big(\chi^{17\lin{7}{10}20}/\epsilon^2\Big)$ & $\O(\chi^{10+p}/\epsilon^2)$\\[-0.6em]
	2D two-step $3\times 3\mapsto 1$, MC & {$\O(\chi^{16}/\epsilon^2)$}  & $\O(\chi^{10}/\epsilon^2)$  & \tq &      $\O\Big(\chi^{16\lin{6}{10}20}/\epsilon^2\Big)$ & $\O(\chi^{10+p}/\epsilon^2)$\\
	2D three-step $3\times 3\mapsto 1$, TL & \dom{$\O(\chi^{12}/\epsilon^2)$}  & $\O(\chi^{8}/\epsilon^2)$  & 8   & \jon{$\O\Big(\chi^{12\lin{5}{8}15}/\epsilon^2\Big)$} & $\O\Big(\chi^{8\lin{2}{8}14}/\epsilon^2\Big)$\\[-0.6em]
	2D three-step $3\times 3\mapsto 1$, TC & \dom{$\O(\chi^{12}/\epsilon^2)$}  & $\O(\chi^{8}/\epsilon^2)$  & \tq &     {$\O\Big(\chi^{12\lin{6}{8}14}/\epsilon^2\Big)$} & $\O\Big(\chi^{7\lin{0}{1}8\lin{2}{8}14/\epsilon^2}\Big)$\\[-0.6em]
	2D three-step $3\times 3\mapsto 1$, MC & \dom{$\O(\chi^{12}/\epsilon^2)$}  & $\O(\chi^{8}/\epsilon^2)$  & \tq &     {$\O\Big(\chi^{12\lin{6}{8}14}/\epsilon^2\Big)$} & $\O\Big(\chi^{7\lin{1}{8}14}/\epsilon^2\Big)$\\
	\hline
	\end{tabular}
	\caption{\label{tab:EMB}\textbf{Variational Monte Carlo with eigenstate measurement basis.} Here, we list the optimal scaling of contraction costs for the VMC approach when employing an eigenstate measurement basis as defined in Eq.~\eqref{eq:EMB2}, where, after every contraction, \emph{all} renormalized sites that leave the causal cone are measured simultaneously (EMB-all). The most costly steps for fMERA and TMERA are indicated in red and blue, respectively. Contraction sequences with the optimal cost scaling are provided in Appendices~\ref{appx:1DbinaryMERA}-\ref{appx:2D3s3x3to1MERA}. The optimized bases lead to increased contraction costs compared to those for VMC with CMB in Tables~\ref{tab:fMERA} and \ref{tab:TMERA}.}
\end{table*}

\section{Eigenstate measurement bases}\label{sec:EMB}
So far, we only considered using the computational measurement basis (CMB) for the VMC algorithms. As described in Ref.~\cite{Ferris2012-85}, one can employ optimized measurement bases, to improve statistical properties during the sampling of energy gradients and other observables. Generally, the optimal choice depends on both the state for the sites to be measured and the observable \cite{Binder2017-95}. As described in Sec.~\ref{sec:fMERA-VMC}, we will employ an eigenstate measurement basis (EMB) [Eq.~\eqref{eq:EMB}]:
Say, we have just contracted a tensor $U_k$, resulting in a causal-cone state $|\psi\ket\in\CC^{\chi^n}\otimes\CC^{\chi^m}$, where we want to measure $n$ of the renormalized sites of the output of $U_k$ that leave the causal cone. We determine the EMB \eqref{eq:EMB} for these $n$ sites by constructing the reduced density operator
\begin{equation}\label{eq:EMB2}
	\varrho_n=\Tr_m|\psi\ket\bra\psi|=\sum_{r=1}^{R}p_r |r\ket\bra r|,
\end{equation}
where $\Tr_m$ denotes the partial trace over the $m$ non-measured sites. In the second step, we have diagonalized $\varrho_n$ yielding the EMB \eqref{eq:EMB} for the $R\leq \chi^n$ nonzero eigenvalues $p_r>0$. The measurement then consists in choosing $r$ according to the probabilities $\{p_r\}$ and projecting the state $|\psi\ket\to (\bra r|\otimes \id_m)|\psi\ket$. As both tasks can be accomplished with one Schmidt decomposition (SVD) of $\psi$ \cite{Nielsen2000}, the associated cost is
\begin{equation}\label{eq:EMB-cost}
	\O(\chi^{\min(2n+m,2m+n)}).
\end{equation}

Note that the EMB of later measurements depend on the results in earlier measurements. To see that this is fine, consider a state $|\psi\ket\in\CC^{\chi^{n_1}}\otimes\CC^{\chi^{n_2}}\otimes\CC^{\chi^{m}}$ in a tripartite Hilbert space, where we measure the first $n_1$ sites in an orthonormal basis $\{|r_1\ket\}$ and denote a fixed orthonormal basis of the following $n_2$ sites by $\{|r_2\ket\}$. For the importance sampling of an expectation value $\bra\psi|\id_{n_1}\otimes\id_{n_2}\otimes h_i|\psi\ket$, the first measurement corresponds to the expansion $\id_{n_1}=\sum_{r_1}|r_1\ket\bra r_1|$ and the second to $\id_{n_2}=\sum_{r_2}|r_2\ket\bra r_2|$. Now, if we choose an alternative $r_1$-dependent basis $\{|r_2(r_1)\ket=U_{r_1}|r_2\ket\}$ for the second measurement, where $U_{r_1}$ is unitary, we still have $\sum_{r_2}|r_2(r_1)\ket\bra r_2(r_1)|=U_{r_1}U_{r_1}^\dag=\id_{n_2}$, i.e., a valid expansion of the identity.

In the following, we consider two different measurement schemes. With ``EMB-all'', we denote the case where, after each tensor contraction in the importance-sampling layer propagation \eqref{eq:VMC-prop} that yields a valid quantum state $|\psi\ket$, the renormalized sites that leave the causal cone are all measured \emph{simultaneously}, i.e., $n$ is maximal in Eq.~\eqref{eq:EMB-cost}. With ``EMB-1'', we denote the case where the sites that leave the causal cone are measured \emph{sequentially}, i.e., $n=1$ in Eq.~\eqref{eq:EMB-cost}.

As the results in Table~\ref{tab:EMB} and Appendices \ref{appx:1DbinaryMERA}-\ref{appx:2D3s3x3to1MERA} show, the additional measurement cost \eqref{eq:EMB-cost} for EMB-all can in fact dominate over the tensor contraction costs and lead to different optimal contraction sequences. For example, the VMC costs for the 1D ternary, 2D nonary two-step, and 2D nonary three-step fMERA increase from $\O(\chi^5/\epsilon^2)$, $\O(\chi^{15}/\epsilon^2)$, and $\O(\chi^{11}/\epsilon^2)$ to
\begin{equation}
	\O(\chi^6/\epsilon^2),\quad
	\O(\chi^{19}/\epsilon^2),\quad\text{and}\quad
	\O(\chi^{12}/\epsilon^2),
\end{equation}
respectively.

With the EMB-1 scheme, the VMC costs for fMERA remain the same as in the CMB scheme (Table~\ref{tab:fMERA}), and the TMERA costs increase more moderately to
\begin{equation*}
	\O\Bigg(\frac{\chi^{5\lin{1}{4}8}}{\epsilon^2}\Bigg),\ \
	\O\Bigg(\frac{\chi^{4\lin{1}{4}7}}{\epsilon^2}\Bigg),\ \ \text{and}\ \
	\O\Bigg(\frac{\chi^{5\lin{1}{4}8}}{\epsilon^2}\Bigg)
\end{equation*}
for the 1D binary, modified binary, and ternary TMERA, respectively, and to order
\begin{equation*}
	\frac{\chi^{15\lin{1}{2}16}}{\epsilon^2},\quad
	\frac{\chi^{15\lin{5}{10}20}}{\epsilon^2},\quad\text{and}\quad
	\frac{\chi^{10\lin{1}{2}11\lin{4}{8}15}}{\epsilon^2}
\end{equation*}
for the 2D quaternary, nonary two-step, and nonary three-step TMERA, respectively.

Given the substantial cost increases, the benefit of EMB-all is questionable. It would be interesting to explore methods that employ improved measurement bases without the need to compute reduced density operators for the measured sites.

\section{Discussion}
By optimizing tensor-network contraction sequences, we have determined the scaling of the MERA computation costs in the bond dimension $\chi$ for four different classical optimization algorithms considering both unconstrained tensors (fMERA) and Trotterized tensors (TMERA), as well as the common EEG optimizations with exact energy gradients and the alternative importance sampling (VMC).

For critical quantum many-body systems, the achievable energy accuracy as well as the optimal number of Trotter steps in TMERA follow power laws $\epsilon\sim\chi^{-\beta}$ and $t\sim\chi^p$, respectively. In combination with the contraction-cost scaling, we obtain algorithmic phase diagrams for each type of MERA. Generally, large $\beta$ (fast energy convergence) favors EEG over VMC, and large $p$ (slow convergence in the number of Trotter steps) favors unconstrained over Trotterized tensors.

We have determined the exponents $\beta$ and $p$ numerically for the 1D modified binary MERA and various critical spin chains. All fall into the EEG region of the algorithmic phase diagram. However, some are close to algorithmic transition lines and it is conceivable that Trotterization and/or VMC may be beneficial for certain models. This may also be influenced by small $\chi$ corrections (subleading cost terms) which we have ignored for simplicity.

Especially for the 2D $2\times 2\mapsto 1$ MERA \cite{Cincio2008-100} and 2D three-step $3\times 3\mapsto 1$ MERA \cite{Evenbly2009-102}, the fMERA VMC and TMERA VMC algorithms prevail in large regions of the algorithmic phase diagram. The results suggest that substantial cost reductions are possible by switching to VMC for 2D systems.
Due to high computation costs, it is however challenging to determine the scaling exponents $\beta$ and $p$ for critical 2D models. Currently, such data is not available, but it could be accessible through MERA VMC simulations. In this light, it appears worthwhile to also reinvigorate the search for 2D MERA structures that reduce the computation cost while the expressiveness of the ansatz is maintained or improved.

In Ref.~\cite{Miao2021_08}, we have discussed a variational quantum algorithm based on TMERA (see also Refs.~\cite{Kim2017_11,Haghshenas2022-12,Miao2023_03,Barthel2023_03,Miao2024-109,Haghshenas2023_05,Job2024_04,Miao2024_12}). Implementing TMERA on quantum computers avoids the high classical tensor contraction costs. Indeed, our numerical analysis in Ref.~\cite{Miao2023_03} suggests a polynomial quantum advantage of quantum TMERA over the classical MERA algorithms already for 1D systems. The separation should be even larger for 2D models.

Another interesting question for future work is whether one could lower classical computation costs further by not insisting on exact representations of the causal-cone density operators in the EEG approach and the causal-cone pure states in the VMC approach. In principle, it could, e.g., be beneficial to employ approximate MPS or PEPS representations or other truncation schemes.

\begin{acknowledgments}
We gratefully acknowledge support from the US National Science Foundation (NSF) Quantum Leap Challenge Institute for Robust Quantum Simulation (award no.\ OMA-2120757).
\end{acknowledgments}

\appendix
\renewcommand{\thesection}{\Alph{section}}
\renewcommand{\thesubsection}{\thesection.\arabic{subsection}}
\makeatletter
\renewcommand{\p@subsection}{}
\renewcommand{\p@subsubsection}{}
\makeatother

\section{Basics on contractions of tensor networks}\label{appx:TNbasics}
Let us prove the basic facts about optimal tensor network contractions as stated in Secs.~\ref{sec:fMERA-EEG} and \ref{sec:fMERA-results}.

\subsection{Contraction of two tensors (Lemma~\ref{lemma1})}\label{appx:lemma1}
To see that in an optimal contraction of two (unconstrained) tensors we can always simultaneously sum over all common indices, consider the contraction task
\begin{equation}\textstyle
	\sum_{i_1=1}^{\chi_1}\sum_{i_2=1}^{\chi_2} A_{i_1,i_2,i_a}B_{i_1,i_2,i_b}
\end{equation}
with $i_a=1,\dotsc,\chi_a$ and $i_b=1,\dotsc,\chi_b$. Executing both sums simultaneously has the cost $\O(\chi_1\chi_2\chi_a\chi_b)$. Executing the sum over $i_1$ first and, then, the one over $i_2$ yields the higher cost $\O(\chi_1\chi_2\chi_a\chi_b+\chi_2\chi_a\chi_b)$.

\subsection{Pairwise contractions only (Lemma~\ref{lemma2})}\label{appx:lemma2}
To see that every network of unconstrained tensors can always be contracted optimally through a sequence of pairwise tensor contractions, consider the three-tensor contraction task
\begin{equation}\textstyle
	\sum_{i_1=1}^{\chi_1}\sum_{i_2=1}^{\chi_2}\sum_{i_3=1}^{\chi_3} A_{i_1,i_2,i_a}B_{i_1,i_3,i_b}C_{i_2,i_3,i_c},
\end{equation}
where $i_a\in[1,\chi_a]$, $i_b\in[1,\chi_b]$, and $i_c\in[1,\chi_c]$ are uncontracted indices.
Executing the three sums simultaneously requires
\begin{equation}\label{eq:tripleContract}
	\sim 2\chi_1\chi_2\chi_3\chi_a\chi_b\chi_c
\end{equation}
multiplications
The pairwise sequential contraction $(AB)C$ requires
\begin{equation}
	\sim \chi_1\chi_2\chi_3\chi_a\chi_b+\chi_2\chi_3\chi_a\chi_b\chi_c
\end{equation}
multiplications, which is never larger than Eq.~\eqref{eq:tripleContract} for any choice of the index dimensions.

\subsection{Disregarding outer products (Lemma~\ref{lemma3})}\label{appx:lemma3}
\begin{figure*}[t]
	\includegraphics[width=0.85\textwidth]{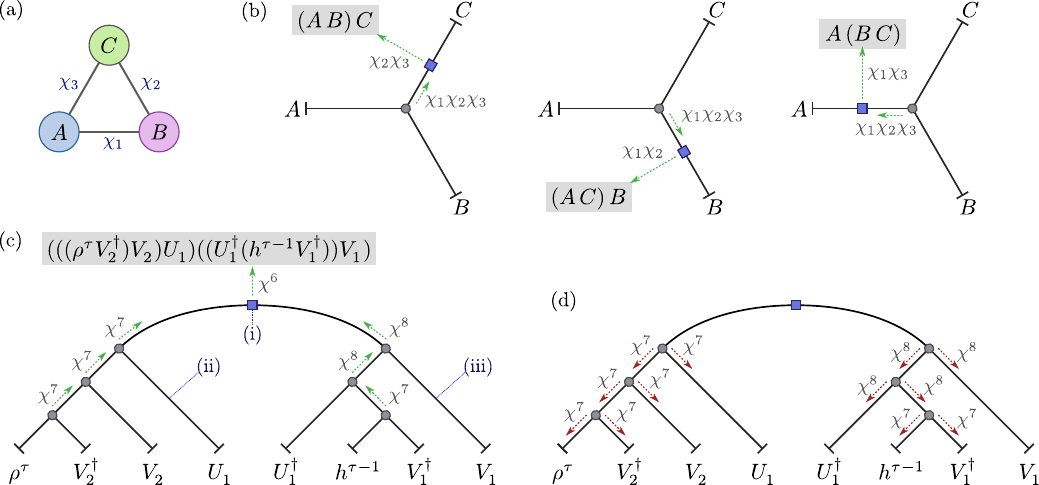}
	\caption{\label{fig:TNcontract}\textbf{Tensor network contraction trees.}
	(a) Graphical representation of the closed three-tensor network \eqref{eq:3tensorNet}. (b) The three possible contraction sequences \eqref{eq:3tensorNetContract} correspond to the three different root positions in the contraction tree. All three options have the leading cost term $\O(\chi_1\chi_2\chi_3)$.
	(c) An optimal contraction sequence $\C$ for the local expectation value $\bra\Psi|\hh_i|\Psi\ket=\Tr\big[\hh_i^{\tau-i}\E_\tL(\dm_i^\tau)\big]$, where the final layer transition $\E_\tL$ is left-moving. The corresponding tensor network is shown in Fig.~\ref{fig:1DternaryMERA}c. The contraction is done by progressing from the leaves (elementary tensors) to the root (blue square). Each pairwise contraction is indicated by an arrow with the associated cost specified next to it.
	(d) In a second pass, we can traverse all vertices with increasing distance from the root, computing the remaining two intermediary tensors for each vertex as indicated by the red arrows.
	The final intermediary tensors along any path are the environment tensors of the corresponding leaf (elementary tensor). Compare, for example, to Eq.~\eqref{eq:1DternaryMERA-envV1} for the $V_1$ environment-tensor contraction sequence.}
\end{figure*}
As described in Sec.~\ref{sec:fMERA-EEG}, for all relevant tensor networks in this work, the tensor dimensions are integer powers $\chi^\nu$ of a base dimension $\chi$. Furthermore, we only care about the scaling $\O(\chi^\alpha)$ of contraction costs, i.e., consider a contraction sequence as \emph{optimal} if its leading cost term has the lowest possible exponent $\alpha$, irrespective of its coefficient and subleading terms.

Without these assumptions, one would need to consider \emph{outer products} in tensor network contractions, i.e., the contraction of two tensors that do not share a common index. As described in Ref.~\cite{Pfeifer2009-79} this requires somewhat intricate case analyses, which we avoid.

To see that, in the case of \emph{unconstrained} tensors, the simplifying assumptions allow us to disregard outer products, consider the contraction task
\begin{equation}\textstyle
	\sum_{i_x=1}^{\chi^x}\sum_{i_y=1}^{\chi^y} A_{i_x,i_a}B_{i_y,i_b}C_{i_x,i_y,i_c},
\end{equation}
where $i_\nu=1,\dotsc,\chi^\nu$ with integer powers $\nu=a,b,c,x,y\geq 1$. The usual contraction $(AC)B$ [and similarly $A(CB)$] has the cost
\begin{equation}\label{eq:no-outer}
	\O(\chi^{a+c+x+y}+\chi^{a+b+c+y}),
\end{equation}
whereas the contraction $(AB)C$, which involves the outer product of $A$ and $B$, has the cost
\begin{equation}
	\O(\chi^{a+b+x+y}+\chi^{a+b+c+x+y}).
\end{equation}
The latter never has a better cost scaling than Eq.~\eqref{eq:no-outer}. Note, however, that optimal contractions of networks with Trotterized tensors may require outer products as discussed in Sec.~\ref{sec:TMERA-costOpt}.

\subsection{Contraction of environment tensors (Lemma~\ref{lemma4})}\label{appx:lemma4}
Consider a tensor network $\T$ consisting of the unconstrained elementary tensors $\{A_1,\dotsc,A_n\}$, Furthermore, let $\T$ be \emph{closed}, i.e., comprise no uncontracted indices. The environment tensor $\partial_{A_k}\T$ is obtained by removing tensor $A_k$ from the network $\T$. Lemma~\ref{appx:lemma4}, which is based on Ref.~\cite{Evenbly2014-89} states that, given an optimal contraction sequence $\C_k$ of cost $\O(\chi^\alpha)$ for the environment of $A_k$, optimal contraction sequences for all other environment tensors $\partial_{A_{k'\neq k}}\T$ can be determined from $\C_k$ and have the same cost.

Let $\C$ be an optimal sequence for the contraction of $\T$, consisting of pairwise contractions (Lemma~\ref{lemma2}). As illustrated in Fig.~\ref{fig:TNcontract}c, each such contraction sequence corresponds to a full binary tree with the $n$ leaves representing the elementary tensors $A_k$, the root with degree two representing the final contraction, and the internal vertices (all having degree three) representing intermediate contraction results $T_i$. Visiting each internal vertex $i$ according to decreasing distance from the root as indicated by the arrows, an intermediary tensor $T_i$ is obtained by contracting the two tensors of its child vertices. According to the \emph{full binary tree theorem}, $\C$ has $n-2$ internal vertices besides the root, implying that we need $n-1$ contractions to evaluate $\T$.

Lemma~\ref{lemma4} is based on the observation that we can move the root through the contraction tree $\C$ without changing the leading contraction-cost term. As illustrated in Fig.~\ref{fig:TNcontract}a, consider the contraction of the closed three-tensor network
\begin{equation}\label{eq:3tensorNet}\textstyle
	\sum_{i_1=1}^{\chi_1}\sum_{i_2=1}^{\chi_2}\sum_{i_3=1}^{\chi_3} A_{i_1,i_3}B_{i_1,i_2}C_{i_2,i_3}
\end{equation}
corresponding to the final two contractions in $\C$. The leading cost term for the three possible contraction sequences is always $\O(\chi_1\chi_2 \chi_3)$;
\begin{subequations}\label{eq:3tensorNetContract}
\begin{align}
	&(A\,B)\,C:\quad\O(\chi_1\chi_2\chi_3 + \chi_2\chi_3),\\
	&(A\,C)\,B:\quad\O(\chi_1\chi_2\chi_3 + \chi_1\chi_2),\\
	&A\,(B\,C):\quad\O(\chi_1\chi_2\chi_3 + \chi_1\chi_3).
\end{align}
\end{subequations}
These three sequences correspond to positioning the root of $\C$ next to $C$, $B$, and $A$, respectively, as indicated in Fig.~\ref{fig:TNcontract}b.
Now, a contraction sequence $\C_k$ for the environment $\partial_{A_k}\T$ for any of the elementary tensors can be obtained from a contraction sequence $\C$ for $\T$ with the root positioned right next to $A_k$ and omitting the final contraction.

Let $\C$ be an optimal sequence for $\T$ with cost $\O(\chi^{\alpha})$. Then, moving the root through the tree to any of the leaves $A_k$ yields an optimal contraction sequence $\C_k$ with cost $\O(\chi^{\alpha})$ for the corresponding environment tensor. Note, that contracting $A_k$ with its environment $\partial_{A_k}\T$ can never be more expensive than building the environment such that omitting the final contraction step never changes the cost scaling. Figure~\ref{fig:TNcontract} shows an example, yielding optimal contraction sequences for the energy environment tensors of the 1D ternary MERA.

Given an optimal $\C$, we can obtain \emph{all} environment tensors $\partial_{A_1}\T,\dotsc,\partial_{A_n}\T$ with a total cost $\O(\chi^{\alpha})$: When moving the root through the tree $\C$, the orientations of the contraction arrows change. For each internal vertex $i$ there are three possible contraction directions, leading to three different intermediary tensors $T_i$ for each vertex. We can obtain all of them by traversing the tree two times. First, pass from the leaves to a root, traversing the edges ordered according to decreasing distance from the root and evaluating the corresponding tensors $T_i$. In Fig.~\ref{fig:TNcontract}, each arrow along an edge indicates the contraction of the tensors coming from the remaining two edges of the vertex. Second, pass from the root to the leaves according to increasing distance from the root and evaluating the remaining two tensors $T_i$ for each vertex as indicated by the red arrows in Fig.~\ref{fig:TNcontract}d. In this way, all possible intermediary tensors (and hence all environments) can be computed with the same cost scaling $\O(\chi^{\alpha})$ as for a single environment contraction.

\begin{widetext}
\section{Optimal contraction sequences for the 1D binary fMERA}\label{appx:1DbinaryMERA}
The 1D binary MERA (Fig.~\ref{fig:MERA-Trotter}a) has a branching ratio $b=2$, causal-cone width $A=3$, and two equivalent layer transitions (left and right-moving).
Example contraction sequences for the left-moving transition of the fMERA with the cost scaling in Tables~\ref{tab:fMERA} and \ref{tab:EMB} are
\begin{gather}\label{eq:opt1Dbin-trans}
   (((((((((\dm\,V^\dag_3)\,V^\pdag_3)\,V^\pdag_1)\,V^\dag_1)\,V^\dag_2)\,U^\dag_1)\,U^\dag_2)\,V^\pdag_2)\,U^\pdag_2)\,U^\pdag_1,\\
   (((((\dm\,V^\pdag_2)\,V^\pdag_3)\,V^\dag_3)\,U^\pdag_2)\,U^\dag_2)\,((((h\,U^\pdag_1)\,U^\dag_1)\,V^\dag_1)\,V^\dag_2)
\end{gather}
for the fMERA layer transition and $V_1$ energy-environment tensor (see Eq.~\eqref{eq:prop-EEG-example} and Fig.~\ref{fig:1DbinaryMERA}a,b),
\begin{equation}\label{eq:opt1Dbin-imp}
   U^\pdag_2\,(U^\pdag_1\,(V^\pdag_2\,(V^\pdag_1\,(V^\pdag_3\,|\psi\ket)))),
\end{equation}
for the fMERA VMC importance sampling with CMB or EMB-all (without explicitly indicating the measurements; see Eq.~\eqref{eq:prop-VMC} and Fig.~\ref{fig:1DbinaryMERA}c),
\begin{gather}
   U^\pdag_1\,(V^\pdag_1\,(U^\pdag_2\,(V^\pdag_3\,(V^\pdag_2\,|\psi\ket)))),\quad
   U^\pdag_1\,(V^\pdag_1\,(U^\pdag_2\,(V^\pdag_3\,(V^\pdag_2\,|\psi\ket)))),\quad
   U^\pdag_1\,((U^\pdag_2\,(V^\pdag_3\,(V^\pdag_2\,|\psi\ket)))\,\bra \psi_h|)
\end{gather}
for the fMERA VMC layer transition of $|\psi\ket$, layer transition of $\bra \psi_h|$, and final environment-tensor contraction with CMB (without explicitly indicating the measurements; see Eq.~\eqref{eq:prop-VMC} and Fig.~\ref{fig:1DbinaryMERA}d), and
\begin{gather}
   U^\pdag_1\,(((\bra r'_1|\,V^\pdag_1)\,(|\psi\ket\,((\bra r'_3|\,V^\pdag_3)\,(\bra r_2|\,U^\pdag_2))))\,V^\pdag_2),\\
   V^\pdag_2\,((\bra r'_3|\,V^\pdag_3)\,((\bra r'_1|\,V^\pdag_1)\,((\bra \psi_h|\,U^\pdag_1)\,(\bra r_2|\,U^\pdag_2)))),\\
   U^\pdag_1\,(((((\bra r'_3|\,V^\pdag_3)\,(\bra r_2|\,U^\pdag_2))\,V^\pdag_2)\,|\psi\ket)\,\bra \psi_h|)
\end{gather}
for the fMERA VMC layer transition of $|\psi\ket$, layer transition of $\bra \psi_h|$, and final environment-tensor contraction with EMB-all.

The sequences \eqref{eq:opt1Dbin-trans} and \eqref{eq:opt1Dbin-imp} are also optimal for the left-moving TMERA EEG layer transition and TMERA VMC importance sampling, respectively, with costs as specified in Tables~\ref{tab:TMERA} and \ref{tab:EMB}.

\section{Optimal contraction sequences for the 1D modified binary fMERA}\label{appx:1DbinaryModMERA}
\begin{figure*}[t]
	\includegraphics[width=0.93\textwidth]{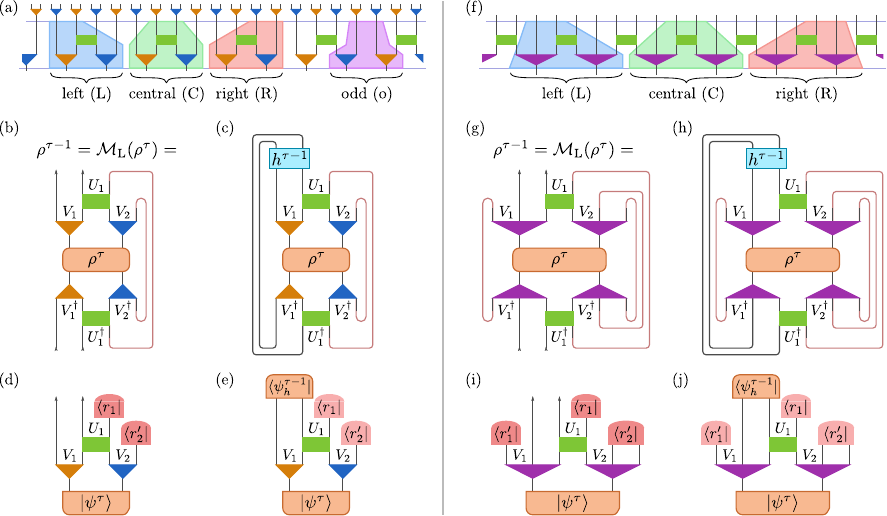}
	\caption{\label{fig:1DbinaryModMERA}\label{fig:1DternaryMERA}\textbf{Tensor networks for 1D modified binary and ternary MERA.}
	The optimization of MERA requires the evaluation of environment tensors.
	(a) The 1D modified binary MERA (Fig.~\ref{fig:MERAs}a), has the shown four different layer transitions. Panels (b-e) show the relevant tensor networks for the (most expensive) left-moving transition to be contracted during the EEG and VMC optimization algorithms. In particular, they show (b) the propagation of causal-cone density operators $\dm^\tau$ in the preparation direction according to Eq.~\eqref{eq:prop-EEG}, (c) the expectation value $\bra\Psi|\hh_i|\Psi\ket=\Tr\big[\hh_i^{\tau-i}\E_\tL(\dm_i^\tau)\big]$, (d) the propagation of pure causal-cone states in the VMC importance sampling as in Eq.~\eqref{eq:prop-VMC}, and (e) the VMC energy-expectation networks from which one obtains gradients (environment tensors) by removing a disentangler or isometry. Optimal contraction sequences are given in Appx.~\ref{appx:1DbinaryModMERA}.
	(f) The 1D ternary MERA (Fig.~\ref{fig:MERAs}b), has the shown three different layer transitions. Panels (g-j) show the relevant tensor networks for the (most expensive) left-moving transition to be contracted during the EEG and VMC optimization algorithms with optimal contraction sequences given in Appx.~\ref{appx:1DternaryMERA}. The lighter color for the projections $\bra r_i|$ in panels (e) and (j) indicate that the measurement results and associated probabilities are already known when the VMC energy(-environment) is being evaluated.}
\end{figure*}
The 1D modified binary MERA (Fig.~\ref{fig:MERAs}a) has branching ratio $b=2$, causal-cone width $A=2$, and four layer transitions -- the equivalent left and right-moving ones (mapping from even to odd bonds), a central one (from even to even bonds), and an odd transition (from odd to even bonds).
Example contraction sequences for the (most expensive) left-moving transition of the fMERA with the cost scaling in Tables~\ref{tab:fMERA} and \ref{tab:EMB} are
\begin{gather}\label{eq:opt1DbinMod-trans}
   (((((\dm\,V^\pdag_2)\,V^\dag_2)\,V^\pdag_1)\,U^\pdag_1)\,V^\dag_1)\,U^\dag_1,\\
   (((((\dm\,V^\dag_2)\,V^\pdag_2)\,V^\dag_1)\,U^\dag_1)\,U^\pdag_1)\,h
\end{gather}
for the fMERA layer transition and $V_1$ energy-environment tensor (see Fig.~\ref{fig:1DbinaryModMERA}b,c),
\begin{equation}\label{eq:opt1DbinMod-imp}
   U^\pdag_1\,(V^\pdag_1\,(V^\pdag_2\,|\psi\ket))
\end{equation}
for the fMERA VMC importance sampling with CMB or EMB-all (without explicitly indicating the measurements; see Fig.~\ref{fig:1DbinaryModMERA}d),
\begin{gather}
   V^\pdag_1\,(U^\pdag_1\,(|\psi\ket\,V^\pdag_2)),\quad
   V^\pdag_1\,(V^\pdag_2\,(U^\pdag_1\,\bra \psi_h|)),\quad
   (V^\pdag_2\,(U^\pdag_1\,\bra \psi_h|))\,|\psi\ket
\end{gather}
for the fMERA VMC layer transition of $|\psi\ket$, layer transition of $\bra \psi_h|$, and final environment-tensor contraction with CMB (without explicitly indicating the measurements; see Fig.~\ref{fig:1DbinaryModMERA}e), and
\begin{gather}
   V^\pdag_1\,(|\psi\ket\,((\bra r'_2|\,V^\pdag_2)\,(\bra r_1|\,U^\pdag_1))),\quad
   V^\pdag_1\,(((\bra r'_2|\,V^\pdag_2)\,(\bra r_1|\,U^\pdag_1))\,\bra \psi_h|),\quad
   (((\bra r'_2|\,V^\pdag_2)\,(\bra r_1|\,U^\pdag_1))\,\bra \psi_h|)\,|\psi\ket
\end{gather}
for the fMERA VMC layer transition of $|\psi\ket$, layer transition of $\bra \psi_h|$, and final environment-tensor contraction with EMB-all.

The sequences \eqref{eq:opt1DbinMod-trans} and \eqref{eq:opt1DbinMod-imp} are also optimal for the left-moving TMERA EEG layer transition and TMERA VMC importance sampling, respectively, with costs as specified in Tables~\ref{tab:TMERA} and \ref{tab:EMB}.

\section{Optimal contraction sequences for the 1D ternary fMERA}\label{appx:1DternaryMERA}
The 1D ternary MERA (Fig.~\ref{fig:MERAs}b) has branching ratio $b=2$, causal-cone width $A=2$, and three layer transitions -- the equivalent left and right-moving ones and a central transition.
Example contraction sequences for the (more expensive) left-moving layer transition of the fMERA with the cost scaling in Tables~\ref{tab:fMERA} and \ref{tab:EMB} are
\begin{gather}\label{eq:opt1Dtern-trans}
   (((((\dm\,V^\pdag_2)\,V^\dag_2)\,U^\pdag_1)\,V^\pdag_1)\,U^\dag_1)\,V^\dag_1,\\
   (((\dm\,V^\dag_2)\,V^\pdag_2)\,U^\pdag_1)\,(U^\dag_1\,(h\,V^\dag_1))
   \label{eq:1DternaryMERA-envV1}
\end{gather}
for the fMERA layer transition and $V_1$ energy-environment tensor (see Fig.~\ref{fig:1DternaryMERA}g,h and Fig.~\ref{fig:TNcontract}c,d),
\begin{equation}\label{eq:opt1Dtern-imp}
   U^\pdag_1\,(V^\pdag_1\,(V^\pdag_2\,|\psi\ket))
\end{equation}
for the fMERA VMC importance sampling  with CMB or EMB-all (without explicitly indicating the measurements; see Fig.~\ref{fig:1DternaryMERA}i),
\begin{gather}
   V^\pdag_1\,(U^\pdag_1\,(V^\pdag_2\,|\psi\ket)),\quad
   V^\pdag_1\,(V^\pdag_2\,(\bra \psi_h|\,U^\pdag_1)),\quad
   (V^\pdag_2\,(\bra \psi_h|\,U^\pdag_1))\,|\psi\ket
\end{gather}
for the fMERA VMC layer transition of $|\psi\ket$, layer transition of $\bra \psi_h|$, and final environment-tensor contraction with CMB (without explicitly indicating the measurements; see Fig.~\ref{fig:1DternaryMERA}j), and
\begin{gather}
   (\bra r'_1|\,V^\pdag_1)\,(((\bra r'_2|\,V^\pdag_2)\,(\bra r_1|\,U^\pdag_1))\,|\psi\ket),\\
   ((\bra \psi_h|\,(\bra r_1|\,U^\pdag_1))\,(\bra r'_1|\,V^\pdag_1))\,(\bra r'_2|\,V^\pdag_2),\\
   (\bra \psi_h|\,((\bra r'_2|\,V^\pdag_2)\,(\bra r_1|\,U^\pdag_1)))\,|\psi\ket
\end{gather}
for the fMERA VMC layer transition of $|\psi\ket$, layer transition of $\bra \psi_h|$, and final environment-tensor contraction with EMB-all.

The sequences \eqref{eq:opt1Dtern-trans} and \eqref{eq:opt1Dtern-imp} are also optimal for the left-moving TMERA EEG layer transition and TMERA VMC importance sampling, respectively, with costs as specified in Tables~\ref{tab:TMERA} and \ref{tab:EMB}.

\section{Optimal contraction sequences for the 2D \texorpdfstring{$2\times 2\mapsto 1$}{2x2-to-1} fMERA}\label{appx:2D2x2to1MERA}
The 2D $2\times 2\mapsto 1$ MERA from Ref.~\cite{Cincio2008-100} has branching ratio $b=4$, causal-cone width $A=3\times 3$, and the four equivalent layer transitions $\E_\TL,\E_\TR,\E_\BL$, and $\E_\BR$. See Figs.~\ref{fig:MERAs}c and \ref{fig:2DMERA-trans}a.
Example contraction sequences for the top-left-moving layer transition $\E_\TL$ of the fMERA with the cost scaling in Tables~\ref{tab:fMERA} and \ref{tab:EMB} are
\begin{gather}\nonumber
   V^\pdag_\ML\,((V^\pdag_\TL\,((((U^\dag_\BL\,((U^\dag_\TR\,((V^\dag_\TC\,((V^\pdag_\MC\,((V^\pdag_\TR\,((V^\pdag_\BC\,((V^\dag_\BL\,((\dm\,((V^\pdag_\MR\,V^\dag_\MR)\,((U^\dag_\BR\,U^\pdag_\BR)\,(V^\pdag_\BR\,V^\dag_\BR))))\\
   \qquad\times V^\pdag_\BL))\,V^\dag_\BC))\,V^\dag_\TR))\,U^\pdag_\TR))\,V^\pdag_\TC))\,V^\dag_\MC))\,U^\pdag_\BL))\,U^\dag_\TL)\,V^\dag_\ML)\,V^\dag_\TL))\,U^\pdag_\TL),
   \label{eq:opt2D2x2to1-trans}\\
   \nonumber
   (((((((((((((((h\,U^\pdag_\TL)\,U^\dag_\TL)\,V^\dag_\TC)\,V^\pdag_\TC)\,U^\dag_\TR)\,V^\dag_\MC)\,U^\pdag_\TR)\,U^\dag_\BL)\,V^\dag_\ML)\,V^\pdag_\ML)\,V^\pdag_\MC)\,U^\pdag_\BL)\,V^\pdag_\BL)\,V^\dag_\BL)\\
   \qquad\times((((((((((\dm\,V^\pdag_\TL)\,V^\dag_\TL)\,V^\dag_\BC)\,V^\pdag_\BC)\,V^\dag_\TR)\,V^\pdag_\TR)\,V^\pdag_\BR)\,V^\dag_\BR)\,V^\dag_\MR)\,V^\pdag_\MR))\,U^\dag_\BR
\end{gather}
for the fMERA layer transition and $U_\BR$ energy-environment tensor,
\begin{equation}\label{eq:opt2D2x2to1-imp}
   U^\pdag_\BL\,(U^\pdag_\TL\,(V^\pdag_\ML\,(U^\pdag_\TR\,(V^\pdag_\TC\,(V^\pdag_\MC\,(U^\pdag_\BR\,(V^\pdag_\TL\,(V^\pdag_\BC\,(V^\pdag_\TR\,(V^\pdag_\BL\,(V^\pdag_\BR\,(V^\pdag_\MR\,|\psi\ket))))))))))))
\end{equation}
for the fMERA VMC importance sampling with CMB or EMB-all (without explicitly indicating the measurements),
\begin{figure*}[t]
	\includegraphics[width=0.84\textwidth]{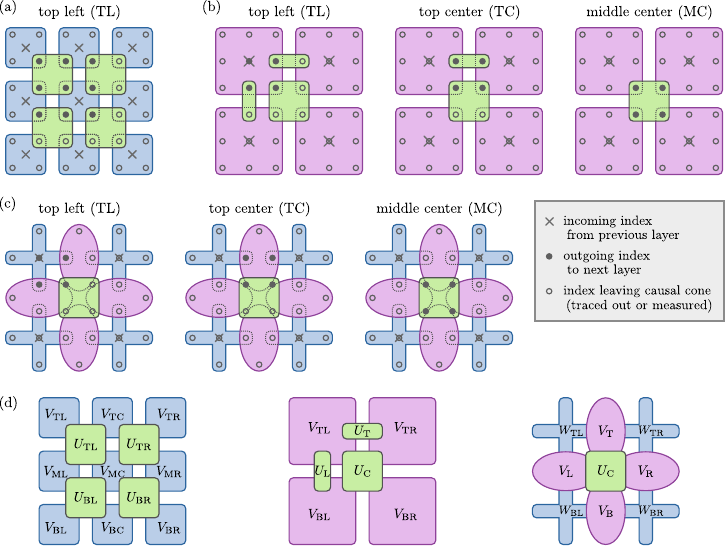}
	\caption{\label{fig:2DMERA-trans}\textbf{Layer transition maps for 2D MERA and labeling.}
	We show representatives for each class of (symmetry-equivalent) layer-transition maps for the (a) 2D $2\times 2\mapsto 1$ MERA, (b) 2D two-step $3\times 3\mapsto 1$ MERA, and (c) 2D three-step $3\times 3\mapsto 1$ MERA from Fig.~\ref{fig:MERAs}c-e. In the EEG approach, sites that leave the causal cone are traced out as in Eq.~\eqref{eq:prop-EEG-example}; in the VMC importance sampling, one applies projective measurements to them as in Eq.~\eqref{eq:prop-VMC}. For a given tensor, all such sites are measured simultaneously and we denote the corresponding measurement basis by $\{|r_\text{X}\ket\}$ for disentanglers $U_\text{X}$, by $\{|r'_\text{X}\ket\}$ for isometries $V_\text{X}$, and by $\{|r''_\text{X}\ket\}$ for disentanglers $W_\text{X}$.}
\end{figure*}
\begin{gather}
   U^\pdag_\TL\,(V^\pdag_\TR\,(V^\pdag_\TC\,(U^\pdag_\TR\,(V^\pdag_\ML\,((V^\pdag_\BL\,((((V^\pdag_\BR\,((V^\pdag_\TL\,|\psi\ket)\,V^\pdag_\MR))\,U^\pdag_\BR)\,V^\pdag_\BC)\,V^\pdag_\MC))\,U^\pdag_\BL))))),\\
   V^\pdag_\TR\,(V^\pdag_\TC\,(U^\pdag_\TR\,((V^\pdag_\BR\,(V^\pdag_\BC\,(V^\pdag_\MR\,((((U^\pdag_\BL\,((U^\pdag_\TL\,V^\pdag_\TL)\,\bra \psi_h|))\,V^\pdag_\ML)\,V^\pdag_\MC)\,U^\pdag_\BR))))\,V^\pdag_\BL))),\\
   V^\pdag_\TR\,(V^\pdag_\TC\,(U^\pdag_\TR\,((\bra \psi_h|\,U^\pdag_\TL)\,(((V^\pdag_\BL\,((V^\pdag_\BC\,(U^\pdag_\BR\,(V^\pdag_\BR\,(V^\pdag_\MR\,|\psi\ket))))\,V^\pdag_\MC))\,U^\pdag_\BL)\,V^\pdag_\ML))))
\end{gather}
for the fMERA VMC layer transition of $|\psi\ket$, layer transition of $\bra \psi_h|$, and final environment-tensor contraction with CMB (without explicitly indicating the measurements), and
\begin{gather}\nonumber
   ((V^\pdag_\BL\,((\bra r'_\ML|\,V^\pdag_\ML)\,(((\bra r_\BL|\,U^\pdag_\BL)\,(((\bra r_\TR|\,U^\pdag_\TR)\,((\bra r'_\TR|\,V^\pdag_\TR)\,(((((\bra r'_\MR|\,V^\pdag_\MR)\\
   \qquad\times((\bra r_\BR|\,U^\pdag_\BR)\,(\bra r'_\BR|\,V^\pdag_\BR)))\,|\psi\ket)\,(\bra r'_\TC|\,V^\pdag_\TC))\,(\bra r'_\BC|\,V^\pdag_\BC))))\,V^\pdag_\MC))\,(\bra r'_\TL|\,V^\pdag_\TL))))\,\bra r'_\BL|)\,U^\pdag_\TL,\\
   \nonumber
   \bra r'_\BL|\,(V^\pdag_\BL\,((V^\pdag_\BR\,((V^\pdag_\MR\,((((((\bra r'_\ML|\,V^\pdag_\ML)\,(((((\bra r_\BL|\,U^\pdag_\BL)\,((\bra r_\TR|\,U^\pdag_\TR)\,(((\bra \psi_h|\,U^\pdag_\TL)\,V^\pdag_\TC)\,\bra r'_\TC|)))\,V^\pdag_\MC)\\
   \qquad\times(\bra r_\BR|\,U^\pdag_\BR))\,(\bra r'_\TR|\,V^\pdag_\TR)))\,\bra r'_\BC|\,V^\pdag_\BC))\,V^\pdag_\TL)\,\bra r'_\TL|))\,\bra r'_\MR|))\,\bra r'_\BR|)),\\
   \nonumber
   \bra r_\BL|\,(\bra r'_\BL|\,(V^\pdag_\BL\,(U^\pdag_\BL\,((((\bra \psi_h|\,U^\pdag_\TL)\,(((\bra r_\TR|\,U^\pdag_\TR)\,((\bra r'_\TR|\,V^\pdag_\TR)\,(((((\bra r'_\MR|\,V^\pdag_\MR)\,((\bra r_\BR|\,U^\pdag_\BR)\\
   \qquad\times(\bra r'_\BR|\,V^\pdag_\BR)))\,|\psi\ket)\,(\bra r'_\TC|\,V^\pdag_\TC))\,(\bra r'_\BC|\,V^\pdag_\BC))))\,V^\pdag_\MC))\,V^\pdag_\ML)\,\bra r'_\ML|))))
\end{gather}
for the fMERA VMC layer transition of $|\psi\ket$, layer transition of $\bra \psi_h|$, and final environment-tensor contraction with EMB-all.

The sequence \eqref{eq:opt2D2x2to1-trans} is also optimal for the top-left-moving TMERA EEG layer transition. Optimal contractions sequences for the layer transitions in the TMERA VMC importance sampling with $t\sim\chi^p$ are
\begin{gather}\label{eq:opt2D2x2to1-TMERAimp-a}
   V^\pdag_\TL\,(U^\pdag_\TL\,(U^\pdag_\BL\,(V^\pdag_\ML\,(U^\pdag_\TR\,(V^\pdag_\TC\,(U^\pdag_\BR\,(V^\pdag_\MC\,(V^\pdag_\BC\,(V^\pdag_\TR\,(V^\pdag_\BL\,(V^\pdag_\BR\,(V^\pdag_\MR\,|\psi\ket))))))))))))
   \ \ \text{for $0<p\leq 2$ and}\\
   \label{eq:opt2D2x2to1-TMERAimp-n}
   U^\pdag_\BL\,(U^\pdag_\TL\,(V^\pdag_\ML\,(U^\pdag_\TR\,(V^\pdag_\TC\,(V^\pdag_\MC\,(U^\pdag_\BR\,(V^\pdag_\TL\,(V^\pdag_\BC\,(V^\pdag_\TR\,(V^\pdag_\BL\,(V^\pdag_\BR\,(V^\pdag_\MR\,|\psi\ket))))))))))))
   \ \ \text{for $2\leq p\leq 8$}
\end{gather}
with costs as specified in Tables~\ref{tab:TMERA} and \ref{tab:EMB}. Note that sequence \eqref{eq:opt2D2x2to1-TMERAimp-n} agrees with Eq.~\eqref{eq:opt2D2x2to1-imp}. Furthermore, note that intermediate contraction results are not always valid quantum states here. Specifically, the isometry $V_\TL$ in sequence \eqref{eq:opt2D2x2to1-TMERAimp-a} is being contracted after the disentangler $U_\TL$.

\section{Optimal contraction sequences for the 2D two-step \texorpdfstring{$3\times 3\mapsto 1$}{3x3-to-1} fMERA}\label{appx:2D2s3x3to1MERA}
The 2D two-step $3\times 3\mapsto 1$ MERA from Ref.~\cite{Evenbly2009-79} has branching ratio $b=9$, causal-cone width $A=2\times 2$, and nine layer transitions which fall into the three inequivalent classes $\{\E_\TL,\E_\TR,\E_\BL,\E_\BR\}$, $\{\E_\TC,\E_\ML,\E_\MR,\E_\BC\}$, and $\{\E_\MC\}$. See Figs.~\ref{fig:MERAs}d and \ref{fig:2DMERA-trans}b.
Example contraction sequences for the top-left-moving layer transition $\E_\TL$ of the fMERA from the most expensive class with the cost scaling in Tables~\ref{tab:fMERA} and \ref{tab:EMB} are
\begin{gather}\label{eq:opt2D2s3x3to1-trans}
   U^\dag_\tL\,(U^\pdag_\tT\,(U^\dag_\tT\,(U^\pdag_\tL\,((\dm\,((V^\pdag_\BL\,V^\dag_\BL)\,((V^\dag_\TR\,V^\pdag_\TR)\,((U^\pdag_\tC\,U^\dag_\tC)\,(V^\pdag_\BR\,V^\dag_\BR)))))\,(V^\dag_\TL\,V^\pdag_\TL))))),\\
   V^\dag_\TL\,(U^\dag_\tL\,(U^\pdag_\tT\,(U^\dag_\tT\,(((\dm\,((V^\pdag_\BL\,V^\dag_\BL)\,((V^\dag_\TR\,V^\pdag_\TR)\,((U^\pdag_\tC\,U^\dag_\tC)\,(V^\pdag_\BR\,V^\dag_\BR)))))\,h)\,U^\pdag_\tL))))
\end{gather}
for the fMERA layer transition and $V_\TL$ energy-environment tensor,
\begin{equation}\label{eq:opt2D2s3x3to1-imp}
   U^\pdag_\tC\,(U^\pdag_\tL\,(U^\pdag_\tT\,(V^\pdag_\TL\,(V^\pdag_\BR\,(V^\pdag_\BL\,(V^\pdag_\TR\,|\psi\ket))))))\quad\text{and}\quad
   (V^\pdag_\BL\,U^\pdag_\tL)\,(U^\pdag_\tC\,((V^\pdag_\TL\,U^\pdag_\tT)\,(V^\pdag_\TR\,(V^\pdag_\BR\,|\psi\ket))))
\end{equation}
for the fMERA VMC importance sampling with CMB and EMB-all (without explicitly indicating the measurements), respectively,
\begin{gather}
   V^\pdag_\TL\,(U^\pdag_\tL\,(V^\pdag_\BL\,(((V^\pdag_\BR\,(V^\pdag_\TR\,U^\pdag_\tC))\,|\psi\ket\,)\,U^\pdag_\tT))),\\
   V^\pdag_\BR\,(V^\pdag_\BL\,(((V^\pdag_\TL\,(U^\pdag_\tL\,(\bra \psi_h|\,U^\pdag_\tT)))\,U^\pdag_\tC)\,V^\pdag_\TR)),\\
   U^\pdag_\tL\,(U^\pdag_\tT\,(\bra \psi_h|\,(V^\pdag_\BL\,((V^\pdag_\BR\,(V^\pdag_\TR\,U^\pdag_\tC))\,|\psi\ket))))
\end{gather}
for the fMERA VMC layer transition of $|\psi\ket$, layer transition of $\bra \psi_h|$, and final environment-tensor contraction with CMB (without explicitly indicating the measurements), and
\begin{gather}
   \bra r_\tL|\,(U^\pdag_\tL\,(((((\bra r'_\TL|\,V^\pdag_\TL)\,(((\bra r'_\BL|\,V^\pdag_\BL)\,((U^\pdag_\tC\,(\bra r'_\TR|\,V^\pdag_\TR))\,\bra r_\tC|))\,|\psi\ket))\,(\bra r'_\BR|\,V^\pdag_\BR))\,U^\pdag_\tT)\,\bra r_\tT|)),\\
   \bra \psi_h|\,(((\bra r_\tL|\,U^\pdag_\tL)\,((\bra r'_\BL|\,V^\pdag_\BL)\,((\bra r_\tC|\,U^\pdag_\tC)\,((((\bra r'_\TL|\,V^\pdag_\TL)\,U^\pdag_\tT)\,(\bra r'_\TR|\,V^\pdag_\TR))\,\bra r_\tT|))))\,(\bra r'_\BR|\,V^\pdag_\BR)),\\
   \bra \psi_h|\,(\bra r_\tL|\,(((((((\bra r'_\BL|\,V^\pdag_\BL)\,((U^\pdag_\tC\,(\bra r'_\TR|\,V^\pdag_\TR))\,\bra r_\tC|))\,(\bra r'_\BR|\,V^\pdag_\BR))\,|\psi\ket)\,U^\pdag_\tT)\,\bra r_\tT|)\,U^\pdag_\tL))
\end{gather}
for the fMERA VMC layer transition of $|\psi\ket$, layer transition of $\bra \psi_h|$, and final environment-tensor contraction with EMB-all.

The sequences \eqref{eq:opt2D2s3x3to1-trans} and \eqref{eq:opt2D2s3x3to1-imp} are also optimal for the top-left-moving TMERA EEG layer transition and TMERA VMC importance sampling, respectively, with costs as specified in Tables~\ref{tab:TMERA} and \ref{tab:EMB}. Furthermore, note that the EMB-all importance-sampling sequence in Eq.~\eqref{eq:opt2D2s3x3to1-imp} involves several postponed measurements due to the intermediate contractions $V_\TL\,U_\tT$ and $V_\BL\,U_\tL$ as well as the contraction of $U_\tC$ coming before the contraction of $V_\BL$.

\section{Optimal contraction sequences for the 2D three-step \texorpdfstring{$3\times 3\mapsto 1$}{3x3-to-1} fMERA}\label{appx:2D3s3x3to1MERA}
The 2D three-step $3\times 3\mapsto 1$ MERA from Ref.~\cite{Evenbly2009-102} has branching ratio $b=9$, causal-cone width $A=2\times 2$, and nine layer transitions which fall into the three inequivalent classes $\{\E_\TL,\E_\TR,\E_\BL,\E_\BR\}$, $\{\E_\TC,\E_\ML,\E_\MR,\E_\BC\}$, and $\{\E_\MC\}$. See Figs.~\ref{fig:MERAs}e and \ref{fig:2DMERA-trans}c.
Example contraction sequences for the top-left-moving layer transition $\E_\TL$ of the fMERA from the most expensive class with the cost scaling in Tables~\ref{tab:fMERA} and \ref{tab:EMB} are
\begin{gather}\label{eq:opt2D3s3x3to1-trans}
   ((W^\dag_\TL\,W^\pdag_\TL)\,(V^\pdag_\tL\,V^\dag_\tL))\,(((U^\dag_\tC\,U^\pdag_\tC)\,((((W^\dag_\TR\,W^\pdag_\TR)\,(V^\pdag_\tR\,V^\dag_\tR))\,((W^\dag_\BR\,W^\pdag_\BR)\,(V^\dag_\tB\,V^\pdag_\tB)))\,((W^\pdag_\BL\,W^\dag_\BL)\,\dm)))\,(V^\pdag_\tT\,V^\dag_\tT)),\\
   W^\dag_\TL\,(V^\pdag_\tT\,((h\,(V^\dag_\tL\,(V^\pdag_\tL\,((((W^\dag_\TR\,W^\pdag_\TR)\,(\dm\,((W^\pdag_\BL\,W^\dag_\BL)\,((W^\pdag_\BR\,W^\dag_\BR)\,(V^\pdag_\tB\,V^\dag_\tB)))))\,(V^\pdag_\tR\,V^\dag_\tR))\,(U^\dag_\tC\,U^\pdag_\tC)))))\,V^\dag_\tT))
\end{gather}
for the fMERA layer transition and $W_\TL$ energy-environment tensor,
\begin{equation}\label{eq:opt2D3s3x3to1-imp}
   U^\pdag_\tC\,(V^\pdag_\tL\,(V^\pdag_\tT\,(W^\pdag_\TL\,(V^\pdag_\tB\,(W^\pdag_\BL\,(V^\pdag_\tR\,(W^\pdag_\TR\,(W^\pdag_\BR\,|\psi\ket))))))))
\end{equation}
for the fMERA VMC importance sampling with CMB or EMB-all (without explicitly indicating the measurements),
\begin{gather}
   V^\pdag_\tL\,(W^\pdag_\TL\,((U^\pdag_\tC\,(W^\pdag_\BL\,((((W^\pdag_\BR\,|\psi\ket)\,V^\pdag_\tR)\,W^\pdag_\TR)\,V^\pdag_\tB)))\,V^\pdag_\tT)),\\
   W^\pdag_\BL\,(W^\pdag_\TR\,((W^\pdag_\BR\,(((V^\pdag_\tL\,(\bra \psi_h|\,(W^\pdag_\TL\,V^\pdag_\tT)))\,U^\pdag_\tC)\,V^\pdag_\tB))\,V^\pdag_\tR)),\\
   \bra \psi_h|\,(V^\pdag_\tL\,((U^\pdag_\tC\,(W^\pdag_\BL\,((((W^\pdag_\BR\,|\psi\ket)\,V^\pdag_\tR)\,W^\pdag_\TR)\,V^\pdag_\tB)))\,V^\pdag_\tT))
\end{gather}
for the fMERA VMC layer transition of $|\psi\ket$, layer transition of $\bra \psi_h|$, and final environment-tensor contraction with CMB (without explicitly indicating the measurements), and
\begin{gather}\nonumber
   (\bra r'_\tT|\,V^\pdag_\tT)\,(((((\bra r'_\tB|\,V^\pdag_\tB)\,((((\bra r''_\TR|\,W^\pdag_\TR)\,((\bra r'_\tR|\,V^\pdag_\tR)\,(\bra r''_\BR|\,W^\pdag_\BR)))\,|\psi\ket)\,(\bra r''_\BL|\,W^\pdag_\BL)))\\
   \qquad\times(\bra r_\tC|\,U^\pdag_\tC))\,(\bra r'_\tL|\,V^\pdag_\tL))\,(\bra r''_\TL|\,W^\pdag_\TL)),\\
   \nonumber
   (\bra r'_\tB|\,V^\pdag_\tB)\,((\bra r''_\BL|\,W^\pdag_\BL)\,((((\bra r'_\tR|\,V^\pdag_\tR)\,(((\bra r'_\tT|\,V^\pdag_\tT)\,(\bra \psi_h|\,((\bra r''_\TL|\,W^\pdag_\TL)\,(\bra r'_\tL|\,V^\pdag_\tL))))\\
   \qquad\times(\bra r_\tC|\,U^\pdag_\tC)))\,(\bra r''_\BR|\,W^\pdag_\BR))\,(\bra r''_\TR|\,W^\pdag_\TR))),\\
   \nonumber
   \bra \psi_h|\,(((((\bra r'_\tB|\,V^\pdag_\tB)\,((((\bra r''_\TR|\,W^\pdag_\TR)\,((\bra r'_\tR|\,V^\pdag_\tR)\,(\bra r''_\BR|\,W^\pdag_\BR)))\,|\psi\ket)\,(\bra r''_\BL|\,W^\pdag_\BL)))\\
   \qquad\times(\bra r_\tC|\,U^\pdag_\tC))\,(\bra r'_\tL|\,V^\pdag_\tL))\,(\bra r'_\tT|\,V^\pdag_\tT))
\end{gather}
for the fMERA VMC layer transition of $|\psi\ket$, layer transition of $\bra \psi_h|$, and final environment-tensor contraction with EMB-all.

The sequences \eqref{eq:opt2D3s3x3to1-trans} and \eqref{eq:opt2D3s3x3to1-imp} are also optimal for the top-left-moving TMERA EEG layer transition and TMERA VMC importance sampling, respectively, with costs as specified in Tables~\ref{tab:TMERA} and \ref{tab:EMB}.

\end{widetext}


\begin{thebibliography}{10}

\bibitem{Karmers1941-60}
H.~A. Kramers and G.~H. Wannier, {\em Statistics of the two-dimensional
  ferromagnet. Part II}, \href{https://doi.org/10.1103/PhysRev.60.263} {Phys.
  Rev. {\bf 60},  263  (1941)}.

\bibitem{Baxter1968-9}
R.~J. Baxter, {\em Dimers on a rectangular lattice},
  \href{https://doi.org/10.1063/1.1664623} {J. Math. Phys. {\bf 9},  650
  (1968)}.

\bibitem{Nightingale1986-33}
M.~P. Nightingale and H.~W.~J. Bl\"ote, {\em Gap of the linear spin-1
  Heisenberg antiferromagnet: A Monte Carlo calculation},
  \href{https://doi.org/10.1103/PhysRevB.33.659} {Phys. Rev. B {\bf 33},  659
  (1986)}.

\bibitem{Fannes1992-144}
M. Fannes, B. Nachtergaele, and R.~F. Werner, {\em Finitely correlated states
  on quantum spin chains}, \href{https://doi.org/10.1007/BF02099178} {Commun.
  Math. Phys. {\bf 144},  443  (1992)}.

\bibitem{White1992-11}
S.~R. White, {\em Density matrix formulation for quantum renormalization
  groups}, \href{https://doi.org/10.1103/PhysRevLett.69.2863} {Phys. Rev. Lett.
  {\bf 69},  2863  (1992)}.

\bibitem{Niggemann1997-104}
H. Niggemann, A. Kl\"umper, and J. Zittartz, {\em Quantum phase transition in
  spin-3/2 systems on the hexagonal lattice - optimum ground state approach},
  \href{https://doi.org/10.1007/s002570050425} {Z. Phys. B {\bf 104},  103
  (1997)}.

\bibitem{Nishino2000-575}
T. Nishino, K. Okunishi, Y. Hieida, N. Maeshima, and Y. Akutsu, {\em
  Self-consistent tensor product variational approximation for 3D classical
  models}, \href{https://doi.org/10.1016/S0550-3213(00)00133-4} {Nucl. Phys. B
  {\bf 575},  504  (2000)}.

\bibitem{Verstraete2004-7}
F. Verstraete and J.~I. Cirac, {\em Renormalization algorithms for quantum-many
  body systems in two and higher dimensions},
  \href{http://arxiv.org/abs/cond-mat/0407066} {arXiv:cond-mat/0407066
  (2004)}.

\bibitem{Vidal-2005-12}
G. Vidal, {\em Entanglement renormalization},
  \href{https://doi.org/10.1103/PhysRevLett.99.220405} {Phys. Rev. Lett. {\bf
  99},  220405  (2007)}.

\bibitem{Schollwoeck2011-326}
U. Schollw\"{o}ck, {\em The density-matrix renormalization group in the age of
  matrix product states}, \href{https://doi.org/10.1016/j.aop.2010.09.012}
  {Ann. Phys. {\bf 326},  96  (2011)}.

\bibitem{Orus2014-349}
R. Or\'{u}s, {\em A practical introduction to tensor networks: Matrix product
  states and projected entangled pair states},
  \href{https://doi.org/10.1016/j.aop.2014.06.013} {Ann. Phys. {\bf 349},  117
   (2014)}.

\bibitem{Cirac2021-65}
J.~I. Cirac, D. P\'erez-Garc\'{\i}a, N. Schuch, and F. Verstraete, {\em Matrix
  product states and projected entangled pair states: Concepts, symmetries, and
  theorems}, \href{https://doi.org/10.1103/RevModPhys.93.045003} {Rev. Mod.
  Phys. {\bf 93},  045003  (2021)}.

\bibitem{Cohen2016-29}
N. Cohen, O. Sharir, and A. Shashua, {\em On the expressive power of deep
  learning: A tensor analysis},
  \href{https://doi.org/10.48550/arXiv.1509.05009} {Ann. Conf. Learn. Theory
  {\bf 29},  698  (2016)}.

\bibitem{Stoudenmire2016-29}
E. Stoudenmire and D.~J. Schwab, {\em Supervised learning with tensor
  networks}, \href{https://doi.org/10.48550/arXiv.1605.05775} {Adv. Neur. Inf.
  Proc. Sys. {\bf 29},  4799  (2016)}.

\bibitem{Novikov2016_05}
A. Novikov, M. Trofimov, and I. Oseledets, {\em Exponential machines},
  \href{https://doi.org/10.48550/arXiv.1605.03795} {International Conference on
  Learning Representations, Workshop Track  (2017)}.

\bibitem{Stoudenmire2018-3}
E.~M. Stoudenmire, {\em Learning relevant features of data with multi-scale
  tensor networks}, \href{https://doi.org/10.1088/2058-9565/aaba1a} {Quantum
  Sci. Technol. {\bf 3},  034003  (2018)}.

\bibitem{Grant2018-4}
E. Grant, M. Benedetti, S. Cao, A. Hallam, J. Lockhart, V. Stojevic, A.~G.
  Green, and S. Severini, {\em Hierarchical quantum classifiers},
  \href{https://doi.org/10.1038/s41534-018-0116-9} {npj Quantum Inf. {\bf 4},
  65  (2018)}.

\bibitem{Liu2019-21}
D. Liu, S.-J. Ran, P. Wittek, C. Peng, R.~B. Garc{\'{\i}}a, G. Su, and M.
  Lewenstein, {\em Machine learning by unitary tensor network of hierarchical
  tree structure}, \href{https://doi.org/10.1088/1367-2630/ab31ef} {New J.
  Phys. {\bf 21},  073059  (2019)}.

\bibitem{Huggins2019-4}
W. Huggins, P. Patil, B. Mitchell, K.~B. Whaley, and E.~M. Stoudenmire, {\em
  Towards quantum machine learning with tensor networks},
  \href{https://doi.org/10.1088/2058-9565/aaea94} {Quantum Sci. Technol. {\bf
  4},  024001  (2019)}.

\bibitem{Cheng2021-103}
S. Cheng, L. Wang, and P. Zhang, {\em Supervised learning with projected
  entangled pair states}, \href{https://doi.org/10.1103/PhysRevB.103.125117}
  {Phys. Rev. B {\bf 103},  125117  (2021)}.

\bibitem{Vieijra2022_02}
T. Vieijra, L. Vanderstraeten, and F. Verstraete, {\em Generative modeling with
  projected entangled-pair states}, \href{http://arxiv.org/abs/2202.08177}
  {arXiv:2202.08177  (2022)}.

\bibitem{Chen2024-46}
H. Chen and T. Barthel, {\em Machine learning with tree tensor networks, CP
  rank constraints, and tensor dropout},
  \href{https://doi.org/10.1109/TPAMI.2024.3396386} {IEEE Trans. Pattern Anal.
  Mach. Intell. {\bf 46},  7825  (2024)}.

\bibitem{Barthel2018-97}
T. Barthel, C. De~Bacco, and S. Franz, {\em A matrix product algorithm for
  stochastic dynamics on networks, applied to non-equilibrium Glauber
  dynamics}, \href{https://doi.org/10.1103/PhysRevE.97.010104} {Phys. Rev. E
  {\bf 97},  010104(R)  (2018)}.

\bibitem{Barthel2020-1}
T. Barthel, {\em The matrix product approximation for the dynamic cavity
  method}, \href{https://doi.org/10.1088/1742-5468/ab5701} {J. Stat. Mech.
  013217  (2020)}.

\bibitem{Crotti2023-120}
S. Crotti and A. Braunstein, {\em Matrix product belief propagation for
  reweighted stochastic dynamics over graphs},
  \href{https://doi.org/10.1073/pnas.2307935120} {Proc. Natl. Acad. Sci. U.S.A.
  {\bf 120},  e2307935120  (2023)}.

\bibitem{Crotti2024_11}
S. Crotti, T. Barthel, and A. Braunstein, {\em Nonequilibrium steady-state
  dynamics of Markov processes on graphs},
  \href{http://arxiv.org/abs/2411.19100} {arXiv:2411.19100  (2024)}.

\bibitem{Vidal2006}
G. Vidal, {\em Class of quantum many-body states that can be efficiently
  simulated}, \href{https://doi.org/10.1103/PhysRevLett.101.110501} {Phys. Rev.
  Lett. {\bf 101},  110501  (2008)}.

\bibitem{Rommer1997}
S. Rommer and S. \"Ostlund, {\em A class of ansatz wave functions for 1D spin
  systems and their relation to DMRG},
  \href{https://doi.org/10.1103/PhysRevB.55.2164} {Phys. Rev. B {\bf 55},  2164
   (1997)}.

\bibitem{PerezGarcia2007-7}
D. Perez-Garcia, F. Verstraete, M.~M. Wolf, and J.~I. Cirac, {\em Matrix
  product state representations},
  \href{https://doi.org/10.48550/arXiv.quant-ph/0608197} {Quantum Info. Comput.
  {\bf 7},  401  (2007)}.

\bibitem{Giovannetti2008-101}
V. Giovannetti, S. Montangero, and R. Fazio, {\em Quantum multiscale
  entanglement renormalization ansatz channels},
  \href{https://doi.org/10.1103/PhysRevLett.101.180503} {Phys. Rev. Lett. {\bf
  101},  180503  (2008)}.

\bibitem{Pfeifer2009-79}
R.~N.~C. Pfeifer, G. Evenbly, and G. Vidal, {\em Entanglement renormalization,
  scale invariance, and quantum criticality},
  \href{https://doi.org/10.1103/PhysRevA.79.040301} {Phys. Rev. A {\bf 79},
  040301(R)  (2009)}.

\bibitem{Evenbly2010-82}
G. Evenbly, P. Corboz, and G. Vidal, {\em Nonlocal scaling operators with
  entanglement renormalization},
  \href{https://doi.org/10.1103/PhysRevB.82.132411} {Phys. Rev. B {\bf 82},
  132411  (2010)}.

\bibitem{Barthel2010-105}
T. Barthel, M. Kliesch, and J. Eisert, {\em Real-space renormalization yields
  finitely correlated states},
  \href{https://doi.org/10.1103/PhysRevLett.105.010502} {Phys. Rev. Lett. {\bf
  105},  010502  (2010)}.

\bibitem{Evenbly2014-112}
G. Evenbly and G. Vidal, {\em Class of highly entangled many-body states that
  can be efficiently simulated},
  \href{https://doi.org/10.1103/PhysRevLett.112.240502} {Phys. Rev. Lett. {\bf
  112},  240502  (2014)}.

\bibitem{Barthel2022-112}
T. Barthel, J. Lu, and G. Friesecke, {\em On the closedness and geometry of
  tensor network state sets}, \href{https://doi.org/10.1007/s11005-022-01552-z}
  {Lett. Math. Phys. {\bf 112},  72  (2022)}.

\bibitem{Landsberg2012-12}
J.~M. Landsberg, Y. Qi, and K. Ye, {\em On the geometry of tensor network
  states}, \href{https://doi.org/10.48550/arXiv.1105.4449} {Quantum Info.
  Comput. {\bf 12},  346  (2012)}.

\bibitem{Kim2017_11}
I.~H. Kim and B. Swingle, {\em Robust entanglement renormalization on a noisy
  quantum computer}, \href{http://arxiv.org/abs/1711.07500} {arXiv:1711.07500
  (2017)}.

\bibitem{Miao2021_08}
Q. Miao and T. Barthel, {\em Quantum-classical eigensolver using multiscale
  entanglement renormalization},
  \href{https://doi.org/10.1103/PhysRevResearch.5.033141} {Phys. Rev. Research
  {\bf 5},  033141  (2023)}.

\bibitem{Haghshenas2022-12}
R. Haghshenas, J. Gray, A.~C. Potter, and G.~K.-L. Chan, {\em Variational power
  of quantum circuit tensor networks},
  \href{https://doi.org/10.1103/PhysRevX.12.011047} {Phys. Rev. X {\bf 12},
  011047  (2022)}.

\bibitem{Miao2023_03}
Q. Miao and T. Barthel, {\em Convergence and quantum advantage of Trotterized
  MERA for strongly-correlated systems}, \href{http://arxiv.org/abs/2303.08910}
  {arXiv:2303.08910  (2023)}.

\bibitem{Barthel2023_03}
T. Barthel and Q. Miao, {\em Absence of barren plateaus and scaling of
  gradients in the energy optimization of isometric tensor network states},
  \href{http://arxiv.org/abs/2304.00161, accepted in Commun.\ Math.\ Phys.}
  {arXiv:2304.00161, accepted in Commun.\ Math.\ Phys.  (2023)}.

\bibitem{Miao2024-109}
Q. Miao and T. Barthel, {\em Isometric tensor network optimization for
  extensive Hamiltonians is free of barren plateaus},
  \href{https://doi.org/10.1103/PhysRevA.109.L050402} {Phys. Rev. A {\bf 109},
  L050402  (2024)}.

\bibitem{Haghshenas2023_05}
R. Haghshenas, E. Chertkov, M. DeCross, T.~M. Gatterman, J.~A. Gerber, K.
  Gilmore, D. Gresh, N. Hewitt, C.~V. Horst, M. Matheny, T. Mengle, B.
  Neyenhuis, D. Hayes, and M. Foss-Feig, {\em Probing critical states of matter
  on a digital quantum computer}, \href{http://arxiv.org/abs/2305.01650}
  {arXiv:2305.01650  (2023)}.

\bibitem{Job2024_04}
J. Job, I.~H. Kim, E. Johnston, and S. Adachi, {\em The cost of entanglement
  renormalization on a fault-tolerant quantum computer},
  \href{http://arxiv.org/abs/2404.10050} {arXiv:2404.10050  (2024)}.

\bibitem{Miao2024_12}
Q. Miao, T. Wang, K.~R. Brown, T. Barthel, and M. Cetina, {\em Probing
  entanglement scaling across a quantum phase transition on a quantum
  computer}, \href{http://arxiv.org/abs/2412.18602} {arXiv:2412.18602  (2024)}.

\bibitem{Ferris2012-85}
A.~J. Ferris and G. Vidal, {\em Variational Monte Carlo with the multiscale
  entanglement renormalization ansatz},
  \href{https://doi.org/10.1103/PhysRevB.85.165147} {Phys. Rev. B {\bf 85},
  165147  (2012)}.

\bibitem{Sandvik2007-99}
A.~W. Sandvik and G. Vidal, {\em Variational quantum Monte Carlo simulations
  with tensor-network states},
  \href{https://doi.org/10.1103/PhysRevLett.99.220602} {Phys. Rev. Lett. {\bf
  99},  220602  (2007)}.

\bibitem{Schuch2008-100}
N. Schuch, M.~M. Wolf, F. Verstraete, and J.~I. Cirac, {\em Simulation of
  quantum many-body systems with strings of operators and Monte Carlo tensor
  contractions}, \href{https://doi.org/10.1103/PhysRevLett.100.040501} {Phys.
  Rev. Lett. {\bf 100},  040501  (2008)}.

\bibitem{Evenbly2009-79}
G. Evenbly and G. Vidal, {\em Algorithms for entanglement renormalization},
  \href{https://doi.org/10.1103/PhysRevB.79.144108} {Phys. Rev. B {\bf 79},
  144108  (2009)}.

\bibitem{Trotter1959}
H.~F. Trotter, {\em On the product of semi-groups of operators},
  \href{https://doi.org/10.1090/S0002-9939-1959-0108732-6} {Proc. Am. Math.
  Soc. {\bf 10},  545  (1959)}.

\bibitem{Suzuki1976-51}
M. Suzuki, {\em Generalized Trotter's formula and systematic approximants of
  exponential operators and inner derivations with applications to many-body
  problems}, \href{https://doi.org/10.1007/BF01609348} {Commun. Math. Phys.
  {\bf 51},  183  (1976)}.

\bibitem{Barthel2020-418}
T. Barthel and Y. Zhang, {\em Optimized Lie-Trotter-Suzuki decompositions for
  two and three non-commuting operators},
  \href{https://doi.org/10.1016/j.aop.2020.168165} {Ann. Phys. {\bf 418},
  168165  (2020)}.

\bibitem{Childs2021-11}
A.~M. Childs, Y. Su, M.~C. Tran, N. Wiebe, and S. Zhu, {\em Theory of Trotter
  error with commutator scaling},
  \href{https://doi.org/10.1103/PhysRevX.11.011020} {Phys. Rev. X {\bf 11},
  011020  (2021)}.

\bibitem{Lam1997-07}
C.-C. Lam, P. Sadayappan, and R. Wenger, {\em On optimizing a class of
  multi-dimensional loops with reduction for parallel execution},
  \href{https://doi.org/10.1142/S0129626497000176} {Parallel Process. Lett.
  {\bf 07},  157  (1997)}.

\bibitem{Hartono2005-155}
A. Hartono, A. Sibiryakov, M. Nooijen, G. Baumgartner, D.~E. Bernholdt, S.
  Hirata, C.-C. Lam, R.~M. Pitzer, J. Ramanujam, and P. Sadayappan,  in {\em
  Computational Science - ICCS 2005}, edited by V.~S. Sunderam, G.~D. van
  Albada, P.~M.~A. Sloot, and J.~J. Dongarra (Springer, Berlin, Heidelberg,
  2005), pp.\ 155--164.

\bibitem{Pfeifer2014-90}
R.~N.~C. Pfeifer, J. Haegeman, and F. Verstraete, {\em Faster identification of
  optimal contraction sequences for tensor networks},
  \href{https://doi.org/10.1103/PhysRevE.90.033315} {Phys. Rev. E {\bf 90},
  033315  (2014)}.

\bibitem{Liang2021-15}
L. Liang, J. Xu, L. Deng, M. Yan, X. Hu, Z. Zhang, G. Li, and Y. Xie, {\em Fast
  search of the optimal contraction sequence in tensor networks},
  \href{https://doi.org/10.1109/JSTSP.2021.3051231} {IEEE J. Sel. Top. Signal
  Process. {\bf 15},  574  (2021)}.

\bibitem{Schindler2020-1}
F. Schindler and A.~S. Jermyn, {\em Algorithms for tensor network contraction
  ordering}, \href{https://doi.org/10.1088/2632-2153/ab94c5} {Mach. Learn.:
  Sci. Technol. {\bf 1},  035001  (2020)}.

\bibitem{Schutski2020-102}
R. Schutski, T. Khakhulin, I. Oseledets, and D. Kolmakov, {\em Simple
  heuristics for efficient parallel tensor contraction and quantum circuit
  simulation}, \href{https://doi.org/10.1103/PhysRevA.102.062614} {Phys. Rev. A
  {\bf 102},  062614  (2020)}.

\bibitem{Jermyn2020-8}
A.~S. Jermyn, {\em Automatic contraction of unstructured tensor networks},
  \href{https://doi.org/10.21468/SciPostPhys.8.1.005} {SciPost Phys. {\bf 8},
  005  (2020)}.

\bibitem{Huang2021-1}
C. Huang, F. Zhang, M. Newman, X. Ni, D. Ding, J. Cai, X. Gao, T. Wang, F. Wu,
  G. Zhang, H.-S. Ku, Z. Tian, J. Wu, H. Xu, H. Yu, B. Yuan, M. Szegedy, Y.
  Shi, H.-H. Zhao, C. Deng, and J. Chen, {\em Efficient parallelization of
  tensor network contraction for simulating quantum computation},
  \href{https://doi.org/10.1038/s43588-021-00119-7} {Nat. Comp. Sci. {\bf 1},
  578?587  (2021)}.

\bibitem{Gray2021-5}
J. Gray and S. Kourtis, {\em Hyper-optimized tensor network contraction},
  \href{https://doi.org/10.22331/q-2021-03-15-410} {Quantum {\bf 5},  410
  (2021)}.

\bibitem{Hauru2021-10}
M. Hauru, M. Van~Damme, and J. Haegeman, {\em Riemannian optimization of
  isometric tensor networks},
  \href{https://doi.org/10.21468/scipostphys.10.2.040} {SciPost Phys. {\bf 10},
   040  (2021)}.

\bibitem{Luchnikov2021-23}
I.~A. Luchnikov, M.~E. Krechetov, and S.~N. Filippov, {\em Riemannian geometry
  and automatic differentiation for optimization problems of quantum physics
  and quantum technologies}, \href{https://doi.org/10.1088/1367-2630/ac0b02}
  {New J. Phys. {\bf 23},  073006  (2021)}.

\bibitem{Evenbly2013}
G. Evenbly and G. Vidal, {\em Quantum criticality with the multi-scale
  entanglement renormalization ansatz}, \href{http://arxiv.org/abs/1109.5334}
  {arXiv:1109.5334  (2011)}.

\bibitem{Cincio2008-100}
L. Cincio, J. Dziarmaga, and M.~M. Rams, {\em Multiscale entanglement
  renormalization ansatz in two dimensions: quantum Ising model},
  \href{https://doi.org/10.1103/PhysRevLett.100.240603} {Phys. Rev. Lett. {\bf
  100},  240603  (2008)}.

\bibitem{Evenbly2009-102}
G. Evenbly and G. Vidal, {\em Entanglement renormalization in two spatial
  dimensions}, \href{https://doi.org/10.1103/PhysRevLett.102.180406} {Phys.
  Rev. Lett. {\bf 102},  180406  (2009)}.

\bibitem{Evenbly2014-89}
G. Evenbly and R.~N.~C. Pfeifer, {\em Improving the efficiency of variational
  tensor network algorithms}, \href{https://doi.org/10.1103/PhysRevB.89.245118}
  {Phys. Rev. B {\bf 89},  245118  (2014)}.

\bibitem{Iten2016-93}
R. Iten, R. Colbeck, I. Kukuljan, J. Home, and M. Christandl, {\em Quantum
  circuits for isometries}, \href{https://doi.org/10.1103/PhysRevA.93.032318}
  {Phys. Rev. A {\bf 93},  032318  (2016)}.

\bibitem{Bethe1931}
H.~A. Bethe, {\em Zur {T}heorie der {M}etalle. I. {E}igenwerte und
  {E}igenfunktionen der linearen {A}tomkette},
  \href{https://doi.org/10.1007/BF01341708} {Z. Phys. {\bf 71},  205  (1931)}.

\bibitem{Hulthen1938}
L. Hulth{\'e}n, {\em {\"U}ber das Austauschproblem eines Kristalles}, {Arkiv
  Mat. Astron. Fys. {\bf 26A},  1  (1938)}.

\bibitem{Cloizeaux1966-7}
J. des Cloizeaux and M. Gaudin, {\em Anisotropic linear magnetic chain},
  \href{https://doi.org/10.1063/1.1705048} {J. Math. Phys. {\bf 7},  1384
  (1966)}.

\bibitem{Johnson1972-6}
J.~D. Johnson and B.~M. McCoy, {\em Low-temperature thermodynamics of the
  $|\Delta|\geq 1$ Heisenberg-ising ring},
  \href{https://doi.org/10.1103/PhysRevA.6.1613} {Phys. Rev. A {\bf 6},  1613
  (1972)}.

\bibitem{Mikeska2004}
H.-J. Mikeska and A.~K. Kolezhuk,  in {\em Quantum Magnetism}, Vol.~645 of {\em
  Lecture Notes in Physics}, edited by U. Schollw\"ock, J. Richter, D.~J.~J.
  Farnell, and R.~F. Bishop (Springer, Berlin, 2004), pp.\ 1--83.

\bibitem{Uimin1970-12}
G.~V. Uimin, {\em One-dimensional problem for $S = 1$ with modified
  antiferromagnetic Hamiltonian}, {JETP Lett. {\bf 12},  225  (1970)}.

\bibitem{Lai1974-15}
C.~K. Lai, {\em Lattice gas with nearest-neighbor interaction in one dimension
  with arbitrary statistics}, \href{https://doi.org/10.1063/1.1666522} {J.
  Math. Phys. {\bf 15},  1675  (1974)}.

\bibitem{Sutherland1975-12}
B. Sutherland, {\em Model for a multicomponent quantum system},
  \href{https://doi.org/10.1103/PhysRevB.12.3795} {Phys. Rev. B {\bf 12},  3795
   (1975)}.

\bibitem{Takhtajan1982-87}
L.~A. Takhtajan, {\em The picture of low-lying excitations in the isotropic
  {H}eisenberg chain of arbitrary spins},
  \href{https://doi.org/10.1016/0375-9601(82)90764-2} {Phys. Lett. A {\bf 87},
  479   (1982)}.

\bibitem{Babujian1982-90}
H. Babujian, {\em Exact solution of the one-dimensional isotropic {H}eisenberg
  chain with arbitrary spins {S}},
  \href{https://doi.org/10.1016/0375-9601(82)90403-0} {Phys. Lett. A {\bf 90},
  479   (1982)}.

\bibitem{Babujian1983-215}
H. Babujian, {\em Exact solution of the isotropic {H}eisenberg chain with
  arbitrary spins: Thermodynamics of the model},
  \href{https://doi.org/10.1016/0550-3213(83)90668-5} {Nucl. Phys. B {\bf 215},
   317   (1983)}.

\bibitem{Laeuchli2006-74}
A. L\"auchli, G. Schmid, and S. Trebst, {\em Spin nematics correlations in
  bilinear-biquadratic $S=1$ spin chains},
  \href{https://doi.org/10.1103/PhysRevB.74.144426} {Phys. Rev. B {\bf 74},
  144426  (2006)}.

\bibitem{Binder2020-102}
M. Binder and T. Barthel, {\em Low-energy physics of isotropic spin-1 chains in
  the critical and Haldane phases},
  \href{https://doi.org/10.1103/PhysRevB.102.014447} {Phys. Rev. B {\bf 102},
  014447  (2020)}.

\bibitem{Alcaraz1988-21}
F.~C. Alcaraz and M.~J. Martins, {\em Conformal invariance and critical
  exponents of the Takhtajan-Babujian models},
  \href{https://doi.org/10.1088/0305-4470/21/23/021} {J. Phys. A: Math. Gen.
  {\bf 21},  4397  (1988)}.

\bibitem{Haldane1983-93}
F.~D.~M. Haldane, {\em Continuum dynamics of the 1-D Heisenberg
  antiferromagnet: Identification with the O(3) nonlinear sigma model},
  \href{https://doi.org/10.1016/0375-9601(83)90631-X} {Phys. Lett. A {\bf 93},
  464   (1983)}.

\bibitem{Haldane1983-50}
F.~D.~M. Haldane, {\em Nonlinear field theory of large-spin Heisenberg
  antiferromagnets: Semiclassically quantized solitons of the one-dimensional
  easy-axis N\'eel state}, \href{https://doi.org/10.1103/PhysRevLett.50.1153}
  {Phys. Rev. Lett. {\bf 50},  1153  (1983)}.

\bibitem{Schulz1986-33}
H.~J. Schulz and T. Ziman, {\em Finite-length calculations of \ensuremath{\eta}
  and phase diagrams of quantum spin chains},
  \href{https://doi.org/10.1103/PhysRevB.33.6545} {Phys. Rev. B {\bf 33},  6545
   (1986)}.

\bibitem{Schulz1986-34}
H.~J. Schulz, {\em Phase diagrams and correlation exponents for quantum spin
  chains of arbitrary spin quantum number},
  \href{https://doi.org/10.1103/PhysRevB.34.6372} {Phys. Rev. B {\bf 34},  6372
   (1986)}.

\bibitem{Affleck1987-36}
I. Affleck and F.~D.~M. Haldane, {\em Critical theory of quantum spin chains},
  \href{https://doi.org/10.1103/PhysRevB.36.5291} {Phys. Rev. B {\bf 36},  5291
   (1987)}.

\bibitem{Alcaraz1992-46}
F.~C. Alcaraz and A. Moreo, {\em Critical behavior of anisotropic spin-S
  Heisenberg chains}, \href{https://doi.org/10.1103/PhysRevB.46.2896} {Phys.
  Rev. B {\bf 46},  2896  (1992)}.

\bibitem{Hallberg1996-76}
K. Hallberg, X.~Q.~G. Wang, P. Horsch, and A. Moreo, {\em Critical behavior of
  the $\mathit{S}\phantom{\rule{0ex}{0ex}}=\phantom{\rule{0ex}{0ex}}3/2$
  antiferromagnetic Heisenberg chain},
  \href{https://doi.org/10.1103/PhysRevLett.76.4955} {Phys. Rev. Lett. {\bf
  76},  4955  (1996)}.

\bibitem{Binder2017-95}
M. Binder and T. Barthel, {\em Symmetric minimally entangled typical thermal
  states for canonical and grand-canonical ensembles},
  \href{https://doi.org/10.1103/PhysRevB.95.195148} {Phys. Rev. B {\bf 95},
  195148  (2017)}.

\bibitem{Nielsen2000}
M.~A. Nielsen and I.~L. Chuang, {\em Quantum Computation and Quantum
  Information} (Cambridge University Press, Cambridge UK, 2000).

\end{thebibliography}
\end{document}